\documentclass[a4paper,11pt]{article}
\pdfoutput=1 

\usepackage{jcappub} 

\usepackage[T1]{fontenc} 

\usepackage{graphicx}
\usepackage{float}
\usepackage{hyperref}
\usepackage{xcolor, xstring}
\usepackage{amssymb, amsmath, mathtools}
\usepackage{listings}
\usepackage{lscape}
\usepackage{lipsum}
\usepackage{multicol}
\usepackage{physics}
\usepackage{natbib}
\usepackage{txfonts}
\usepackage[small]{caption}
\usepackage{xspace}
\usepackage{comment}
\usepackage[normalem]{ulem} 

\newcommand{\pot}[2]{#1 \times 10^{#2}}

\newcommand{\boost}{x\partial_x}
\newcommand{\diffusion}{\mathcal{D}_x}
\newcommand{\diffusionStar}{\mathcal{D}^*_x}
\newcommand{\sterling}[2]{\genfrac{[}{]}{0pt}{0}{#1}{#2}}
\newcommand{\eulerian}[2]{\genfrac{\langle}{\rangle}{0pt}{0}{#1}{#2}}
\newcommand{\boostO}{\hat{\mathcal{O}}_x}

\newcommand{\KompO}{\hat{\mathcal{K}}_x}

\newcommand{\DiffO}{\hat{\mathcal{D}}_x}
\newcommand{\DiffstarO}{\hat{\mathcal{D}}^*_x}

\newcommand{\nbb}{n_{\rm bb}}
\newcommand{\Gspec}{{G}}
\newcommand{\Yspec}{{{Y}}}
\newcommand{\Ynspec}[1]{{Y}_{#1}}
\newcommand{\Mspec}{{M}}

\newcommand{\zf}{z_{\rm f}}

\newcommand{\COBEF}{{\it COBE/FIRAS}\xspace}

\newcommand{\GHz}{{\rm GHz}}

\newcommand{\expf}[1]{{{\rm e}^{#1}}}

\newcommand{\Tz}{{T_{z}}}

\newcommand{\zh}{{z_{\rm h}}}
\newcommand{\zmu}{{z_{\mu}}}

\newcommand{\xc}{x_{\rm c}}

\newcommand{\id}{{\,\rm d}}

\newcommand{\beq}{\begin{equation}}   %

\newcommand{\eeq}{\end{equation}}   %

\newcommand{\beqa}{\begin{eqnarray}}   %

\newcommand{\eeqa}{\end{eqnarray}}   %

\newcommand{\bealf}[1]{\begin{align} #1 \end{align}}

\newcommand{\beal}{\begin{align}}
\newcommand{\enal}{\end{align}}

\newcommand{\bspl}{\begin{split}}

\newcommand{\espl}{\end{split}}

\newcommand{\bsub}{\begin{subequations}}

\newcommand{\esub}{\end{subequations}}

\newcommand{\bmulti}{\begin{multline}}   %

\newcommand{\beqm}{\begin{mathletters}}   %

\newcommand{\eeqm}{\end{mathletters}}   %

\newcommand{\me}{m_{\rm e}}

\newcommand{\Ne}{N_{\rm e}}

\newcommand{\Te}{T_{\rm e}}

\newcommand{\Tg}{T_{\gamma}}

\newcommand{\The}{\theta_{\rm e}}

\newcommand{\sigT}{\sigma_{\rm T}}

\newcommand{\vek} [1]{\mbox{\boldmath${#1}$\unboldmath}}

\newcommand{\Thz}{\theta_{z}}

\newcommand{\ysc}{y_{z}}

\usepackage{bbm}

\title{Spectro-spatial evolution of the CMB I: discretisation of the thermalisation Green's function}

\author[a]{Jens Chluba}

\author[a]{, Thomas Kite}

\author[b,c]{and Andrea Ravenni}

\affiliation[a]{Jodrell Bank Centre for Astrophysics, School of Physics and Astronomy, The University of Manchester, Oxford Road, Manchester, M13 9PL, U.K.}

\affiliation[b]{Dipartimento di Fisica e Astronomia ``Galileo Galilei'', Università degli Studi di Padova, via Marzolo 8, I-35131, Padova, Italy.}
\affiliation[c]{INFN, Sezione di Padova, via Marzolo 8, I-35131, Padova, Italy.}

\emailAdd{Jens.Chluba@Manchester.ac.uk}
\emailAdd{Thomas.Kite@Manchester.ac.uk}
\emailAdd{Andrea.Ravenni@unipd.it}

\date{Aug 2022}

\begin{document}

\abstract{Spectral distortions of the cosmic microwave background (CMB) have been recognized as an important future probe of the early Universe. Existing theoretical studies primarily focused on describing the evolution and creation of {\it average distortions}, ignoring spatial perturbations in the plasma. One of the main reasons for this choice is that a treatment of the spectro-spatial evolution of the photon field deep into the primordial Universe requires solving a radiative transfer problem for the distortion signals, which in full detail is computationally challenging. 
Here we provide the first crucial step towards tackling this problem by formulating a new spectral discretisation of the underlying average thermalisation Green's function. 
Our approach allows us to convert the high-dimensional partial differential equation system ($\simeq 10^3-10^4$ equations) into and set of ordinary differential equations of much lower dimension ($\simeq 10$ equations).
We demonstrate the precision of the approach and highlight how it may be further improved in the future. 
We also clarify the link of the observable spectral distortion parameters (e.g., $\mu$ and $y$) to the computational spectral basis that we use in our frequency discretisation. This reveals how several basis-dependent ambiguities can be interpreted in future CMB analysis.
Even if not exact, the new Green's function discretisation can be used to formulate a generalised photon Boltzmann-hierarchy, which can then be solved with methods that are familiar from theoretical studies of the CMB temperature and polarisation anisotropies. We will carry this program out in a series of companion papers, thereby opening the path to full spectro-spatial exploration of the CMB with future CMB imagers and spectrometers.}

\maketitle
\flushbottom

\newpage
\section{Introduction}
Spectral distortions (SDs) of the cosmic microwave background (CMB) have now been recognized as an important future probe of early-universe and particle physics. In particular the ability of CMB SDs to constrain the primordial power spectrum at small scales \citep{Sunyaev1970mu, Daly1991, Hu1994, Chluba:2x2} provides important motivation to push the observational frontier with the next generation CMB experiments~\citep{Delabrouille2021, Chluba2021ExA}. However, much of the recent theoretical work \citep[e.g.,][]{Chluba2011therm, Sunyaev2013, Tashiro2014, Lucca2020, Chluba2020large} and experimental spectrometer concept studies \citep{Kogut2011PIXIE, PRISM2013WPII, Kogut2016SPIE, Kogut2019BAAS, BISOU2021, Cosmo2021} focused primarily on the science of {\it average distortion signals}. While from the theoretical point of view it is clear that distortion anisotropies should be smaller and thus harder to detect, one of the main reasons for this preference is that the computations of distortion anisotropies of primordial origin are difficult and currently beyond the possibilities of existing Boltzmann codes. 

To illustrate this statement we highlight that numerically solving the full thermalisation problem for SDs created on average by energy release now takes of order $\simeq 30~{\rm seconds}$ on a standard laptop using {\tt CosmoTherm}~\citep{Chluba2011therm, Acharya2021large}. 
While this is already highly optimised, it will be difficult to extend this method to SD anisotropies, where in analogy to the standard CMB temperature fluctuations \citep{CLASSCODE, CAMB} one would have to solve the thermalisation problem for multiple $k$-modes. For each $k$-mode, a multipole hierarchy would furthermore be required, overall boosting the computation by a factor $\simeq 10^3$. In addition, one would have to consider how to convert the final (frequency-dependent) signal transfer functions into CMB observables, which further increases the complexity of the problem over the standard CMB anisotropy computation, likely yielding single computations that would take $\mathcal{O}(10^5-10^6)\,{\rm seconds}$. While not necessarily prohibitively expensive with modern computational resources, this brute force approach would be overly-complicated for exploratory calculations and not scalable in parameter forecasts and searches for new physics.

{\it How could one make the problem more tractable?} The most common approach is to simplify the problem by considering limiting cases.
In particular, scenarios in which the evolution of distortions and primordial perturbations as well as thermalisation physics can be mostly separated come to mind. 
This brings us to the well-known Sunyaev-Zeldovich (SZ) effect \citep{Zeldovich1969, Sunyaev1980}, which is created by anisotropic heating effects in the late Universe, sourcing $y$-type distortion anisotropies that peak at several arcminute angular scales. This signal is highly non-Gaussian and requires an understanding of the non-linear large-scale structure evolution, but then analytically translates the statistical properties of the dark matter distribution into the $y$-field \citep{Komatsu1999, Refregier2000, Komatsu2002, Hill2013, Bolliet:2017lha}. The SZ effect is therefore an important probe for cosmology and cluster physics \citep{Carlstrom2002, SZreview2019}.

Another example is the sourcing of $y$-distortion anisotropies by the mixing of blackbodies in the perturbed universe \citep{Sunyaev1970mu, Daly1991}. This second order effect leads to a fluctuating $y$-distortion sky \citep{Stebbins2007, Pitrou2010} in addition to an average distortion \citep{Chluba:2x2} when perturbations dissipate by free-streaming and Thomson scattering effects. For the fluctuating part, no spectral evolution has to be considered at the late stages (redshift $z\lesssim 10^4$), just like for the SZ effect -- a linear perturbation description of the problem is furthermore possible, yielding $y$-parameter transfer functions that are excited by first order temperature perturbations \citep{Pitrou2010, Ota2017}. If the amplitude of the small-scale curvature perturbations is modulated by large-scale modes this can furthermore lead to correlated $\mu\times T$ and $y\times T$ fluctuations \citep{Pajer2012, Ganc2012, Biagetti2013, Razieh2015, Ota2016, Chluba2017muT, Ravenni2017, Haga2018}, which can be directly constrained using CMB imagers \citep[see][for most recent forecasts and constraints]{Remazeilles2022, Rotti2022, Bianchini:2022dqh}. Note that at the largest angular scales, the corresponding transfer problem was simplified by neglecting details of the distortion evolution in the perturbed Universe \citep{Pajer2012, Pajer2012b, Chluba2017muT}.

There are, however, a number of aspect to the thermalisation problem that have not been captured by any of these calculations. As explained in \cite{Chluba:2x2}, if an average distortion is present during the pre-recombination era, the standard density perturbations at first order will source distortion anisotropies. Assuming the average SD is $\Delta n^{(0)}_\nu$ in terms of the photon occupation number, the SD anisotropies will have a spectrum that follows $\Delta n^{(1)}_\nu\propto -\nu \partial_\nu \Delta n^{(0)}_\nu$ \citep{Chluba:2x2}. Even without any spectral evolution, the standard Doppler terms and potential perturbations therefore source distortion anisotropies, which have not been evaluated. Assuming that the average distortion saturates the limits imposed by \COBEF \citep{Mather1994, Fixsen1996}, one can expect distortion anisotropies at the level of $\simeq 10^{-8}$--$10^{-7}$ of the average CMB. This can exceed the signals expected from the aforementioned non-Gaussian signals and can also be directly constrained with existing and future CMB imaging data.
In addition, the thermalisation efficiency should vary from patch to patch in the perturbed Universe. The required terms in the photon Boltzmann equation were already discussed in \citep{Chluba:2x2}; however, only recently has the effect been estimated using a {\it separate universe} approach \citep{Zegeye2022}. In particular for modes that cross the horizon at or after the recombination process completes this effect should be noticeable in the transfer function solutions, but has not been computed using a full Boltzmann treatment.

To fully capitalise on the potential of spectral distortion anisotropy studies, we need to formulate a generalized photon Boltzmann equation that goes beyond the standard temperature and polarisation anisotropies. The biggest bottleneck is due to the discretisation of the spectral evolution, which currently is done with $\simeq 10^3$--$10^4$ bins in frequency, as explained above.
In this work, we obtain a new discretisation for the average frequency evolution that reduces the computational burden by a factor of $\simeq 10^3$  (Sect.~\ref{sec:ODE_Greens}). This allows us to model the thermalisation from $y\rightarrow\mu\rightarrow T$ with a small number ($\simeq 10$) of new spectral parameters, that can represent the exact calculation from {\tt CosmoTherm} to high precision. 
In contrast to other approximations, the solution is no longer limited to the three standard spectral shapes but allows one to capture the dominant contributions from the residual distortion \citep[e.g.,][]{Chluba2013PCA}. We also explain how the computational distortion parametrisation can be mapped back onto the leading residual distortion spectra, which present the main spectral shapes that may be testable in future applications (Sect.~\ref{sec:observables}). 

This paper is the first in a series of works that study the effect of spectro-spatial evolution of the CMB. In paper II (in preparation) we will formulate the generalised Boltzmann equation, strongly drawing on the results of this paper. 
In paper III (in preparation), we will present a detailed discussion of the distortion transfer functions and power spectra, highlighting the importance of various physical effects and providing Fisher forecasts. We also plan subsequent papers that discuss how the dissipation of acoustic modes in the presence of primordial non-Gaussianty causes spectral distortion anisotropies, and which constraints on various scenarios can be expected. Overall we hope this will provide further motivation to study SDs in the future.

\section{Approximate ODE representation of the thermalisation Green's function}
\label{sec:ODE_Greens}
In this section, we establish a novel way of modeling the spectral evolution of the average photon field under repeated Compton scattering and thermal photon emission processes. In terms of perturbation theory, this is akin to focusing on the background quantities only, which leads back to the Green's function approach for the thermalisation problem, as will be develope here.

\subsection{Brief recap of the thermalisation Green's function}
\label{sec:recap}
The efficiency of photon production and Comptonisation in the primordial plasma dictate various eras with characteristic SD shapes which are defined below. At sufficiently early times, in the temperature or $T$-era ($\pot{2}{6} \lesssim z$), thermalisation processes are very efficient and any excess energy is rapidly converted into a temperature shift, $\Gspec(x)$. Here $x=h\nu/k \Tz$ where $\Tz=T_0(1+z)$ is the background reference temperature, which is chosen to match today's CMB temperature $T_0=2.7255\,{\rm K}$ \citep{Fixsen2009}.\footnote{This avoids having to deal with redshifting terms.} 
The subsequent $\mu$-era ($\pot{5}{4} \lesssim z \lesssim \pot{2}{6}$) is characterised by a lack of photon production, leading to a chemical potential distortion, $\Mspec(x)$. Finally, the $y$-era ($z \lesssim \pot{5}{4}$) renders photon energy redistribution inefficient, leading to a distortion, $\Yspec(x)$, related to the well-known SZ effect, albeit in this case of primordial origin.

The different characteristic spectra introduced above have the forms \citep[e.g.,][]{Chluba2018, Lucca2020}
\begin{equation}
\label{eq:basic_spectra}
\Gspec(x)
=\frac{x\,\expf{x}}{(\expf{x}-1)^2},
\qquad 
\Yspec(x)
=\Gspec(x)\left[x\frac{\expf{x}+1}{\expf{x}-1}-4\right],
\qquad 
\Mspec(x)=\Gspec(x)\left[\frac{1}{\beta_M}-\frac{1}{x}\right],
\end{equation}
with $\beta_M=3\zeta(3)/\zeta(2)\approx 2.1923$. In Sect.~\ref{sec:SED_basis} we will describe how these can be obtained by boosts of the average blackbody spectrum. Central properties of these spectra are summarised by their dimensionless photon number density $N_f=\int x^2 f(x) \id x$ and energy density $E_f=\int x^3 f(x) \id x$. The corresponding integrals can be carried out analytically in terms of Riemann $\zeta$-functions. Also using the blackbody occupation number,  $\nbb(x)=1/(\expf{x}-1)$, we then find:
\bsub
\label{eq:basic_spectra_moments}
\begin{align}
N_{\nbb} &= 2\zeta(3) \approx 2.40411, 
& E_{\nbb} &= \frac{\pi^4}{15} \approx 6.49394
\\
N_{\Gspec} &= 6\zeta(3) =3N_{\nbb} \approx 7.21234,
& E_{\Gspec} &= E_{\Yspec} = \frac{4\pi^4}{15} = 4E_{\nbb} \approx 25.9758
\\
N_{\Yspec} &= N_{\Mspec} = 0, 
& E_{\Mspec} &= \frac{2 \pi^6}{135 \zeta(3)} - 6\zeta(3) \approx \frac{E_{\nbb}}{1.40066} \approx 4.63635.
\end{align}
\esub
The absence of overall photon number for $y$ and  $\mu$ type distortions is by construction (and easily achieved by subtracting $\Gspec$ from alternative definitions). This convention has already been commonplace in the literature, but will become a fundamental simplifying fact in the novel treatment introduced below.

While the heuristic decomposition into three distinct eras introduced above conveys the correct physics to relatively high precision, it is much more convenient to have a robust framework in which the results can be expanded and built upon. In \cite{Chluba2013Green} it was shown that the thermalisation problem can be expressed as a Green's function problem in the limit of small energy injection:
\begin{equation}
    s = \alpha_f\int_0^\infty \mathcal{J}_s(z) \,\frac{\dd \mathcal{Q}}{\dd z}\,\id z,
\end{equation}
where $s\in\{\Theta\equiv \Delta T/\Tz, \mu, y\}$ gives the signal amplitude of the corresponding SD ($f\in\{\Gspec, \Yspec, \Mspec\}$), $\mathcal{J}_s$ is a dimensionless energy branching ratio, and $\alpha_f\equiv E_{\nbb}/E_f\in\{1/4, 1/4, 1.40066\}$ is an energetic conversion factor from a blackbody spectrum to the SD amplitude [easily read off from Eq.~\eqref{eq:basic_spectra_moments}].
The energy release is determined by the comoving relative energy injection rate, $\frac{\dd \mathcal{Q}}{\dd z}= \frac{1}{\rho_\gamma}\frac{\dd Q_{\rm c}}{\dd z}$, where $\dd Q_{\rm c}/\dd z$ directly follows from the photon collision term.

For clarity we note that the \textit{three era} picture of the early Universe would correspond to simple top-hat functions for $\mathcal{J}_s$ \citep[i.e., see `Method A' in][]{Chluba2016}. Other approximations for the energy branching ratios of varying accuracy exist \citep{Chluba2016}, including the addition of intermediate spectral shapes known as {\it residual distortions} \citep{Chluba2013PCA} and perturbative SD approximations for moderate scattering $y$-parameter~\citep{Khatri2012mix}. A byproduct of this work is the ability to generate accurate Green's functions in a generalized spectral basis to a precision comparable with full numerical treatments (see Sect.~\ref{sec:basic_idea}).

While the spectral shapes in Eq.~\eqref{eq:basic_spectra} are physically motivated -- each characteristic of a limiting case for each phase in the early Universe -- they are insufficient to model the general evolution of the spectrum. In the following sections we introduce a method for extending this set of spectral functions and explain how this new spectral basis eventually allows for full spectro-spatial solutions of primordial perturbations in the photon field.

\subsection{Basic idea and lowest order solution of the thermalisation problem}
\label{sec:basic_idea}
As previously mentioned, in the $\mu$-era all injected energy rapidly converts into a the $\mu$-distortion \citep{Sunyaev1970mu, Danese1982, Burigana1991, Hu1993}. The net $\mu$-parameter is given by the evolution equation
$\frac{\partial \mu}{\partial t}
\approx\gamma_\rho\,\dot{\mathcal{Q}}$,
where $\gamma_\rho\equiv \alpha_\Mspec\approx 1.4007$ and $\dot{\mathcal{Q}}=\dd\mathcal{Q}/\dd t$.
For a given $\dot{\mathcal{Q}}$, this equation can be solved with initial $\mu=0$.

\begin{figure}
\centering
\includegraphics[width=\columnwidth]{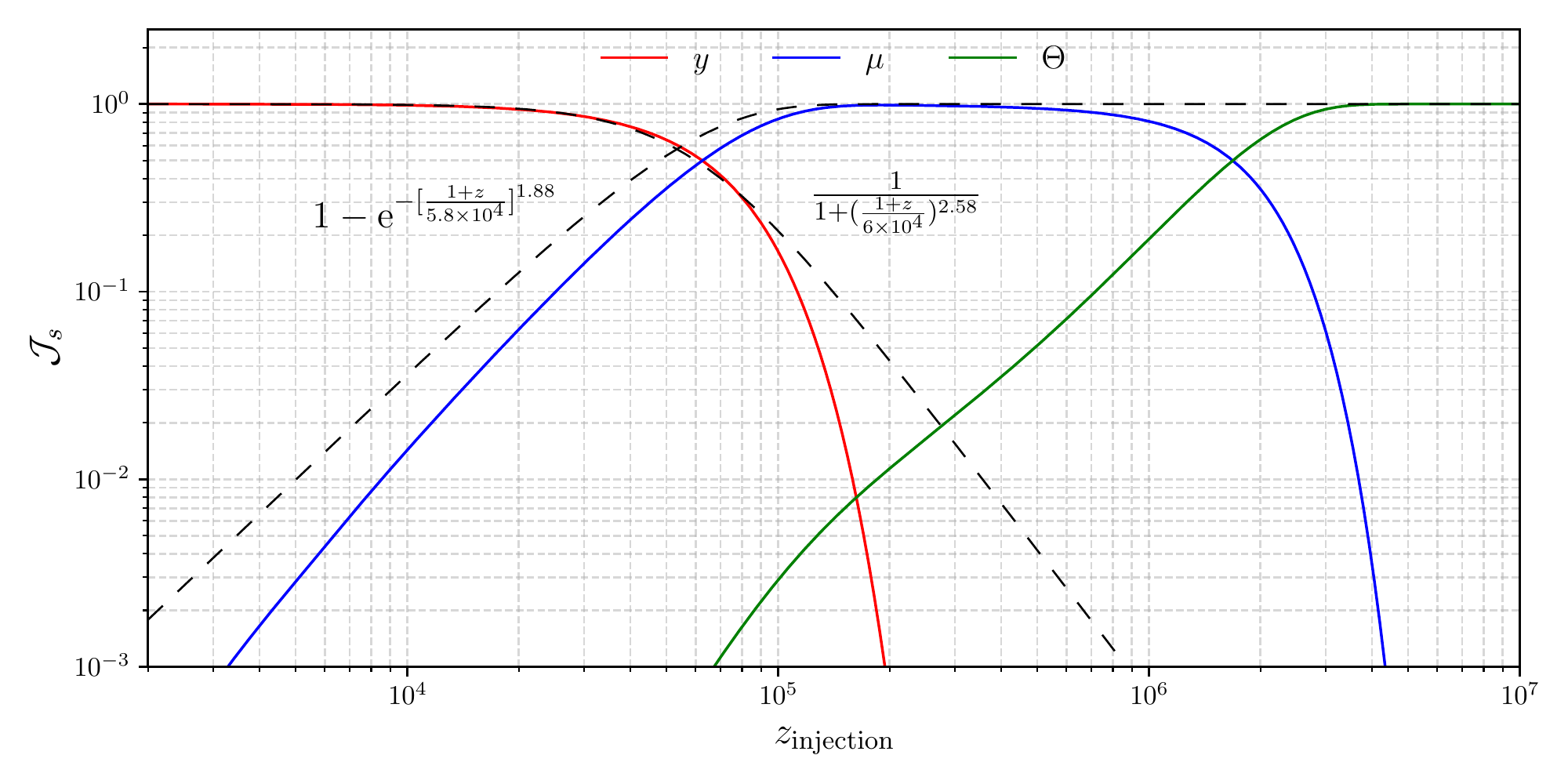}
\\
\caption{Fractions of energy in $y$, $\mu$ and $\theta$ as seen today after injecting a normalised narrow Gaussian of energy at redshift $z_{\rm injection}$. This illustrates that the Green's function, as determined by Eq.~\eqref{eq:simple_ymuT}, already broadly reproduces the least square fit results given in \citep{Chluba2013Green} based on the frequency-binned Green's function. The transition redshift, $z_{y\mu}\approx \pot{5}{4}$, to the $\mu$-regime is obtained in both approximations. However, in comparison to the least square fits, $\mathcal{J}_y$ decays more rapidly towards high redshift.}
\label{fig:Greens_comparison}
\end{figure}

Physically, the energy injection first leads to an increase in the distortion $y$-parameter by $\dot y \approx \frac{1}{4}\dot{\mathcal{Q}}$, which then quickly converts into $\mu$. If we insert this intermediate step, we may instead write
\bealf{
\label{eq:simple_ymu}
\frac{\partial y}{\partial t}
&\approx \frac{1}{4}\dot{\mathcal{Q}}- 4\dot\tau \Thz\,y
\qquad\text{and}\qquad 
\frac{\partial \mu}{\partial t}
\approx\gamma_\rho\,4\dot\tau \Thz\,(4y)\approx 22.411\,\dot\tau \Thz\,y.
}
Here, $4y$ is the relative momentary energy density within the $y$-distortion part, $\Thz=k_{\rm B} \Tz/\me c^2$ is the dimensionless temperature,\footnote{We will use dimensionless temperatures, $\theta_X=k T_X/\me c^2$, frequently, with $T_X\in \{\Te, \Tz, \Tg\}$.} and $\dot\tau=\id\tau/\id t=\Ne\,\sigT c$ denotes the  differential Thomson optical depth, all with the common choice of constants. The average energy exchange rate is $\langle\Delta\nu/\nu\rangle \simeq 4\Thz$ per scattering \citep{Sazonov2001, CSpack2019}, which determines how quickly energy flows from $y$ to $\mu$.
This identifies $\dot{\tau}\Thz$ as a fundamental timescale in the thermalisation problem which contrasts with the timescale of Thomson scattering $\dot{\tau}$ -- a fact which will become important for the generalised Boltzmann hierarchy (paper~II). 
As we see in Fig.~\ref{fig:Greens_comparison}, the solution of this simple system roughly captures the transition between the $\mu$ and $y$-eras, yielding a $y$-distortion visibility $\mathcal{J}_y\approx \expf{-4\,\ysc}$, and, from energy conservation, the $\mu$-visibility $\mathcal{J}_\mu\approx 1-\mathcal{J}_y$.

Following \citep{Chluba2014}, the reduction of the chemical potential by the Bremsstrahlung (BR) and double Compton (DC) processes is approximately given by $\dot\mu|_{\rm em/abs}\approx-\gamma_N\,\xc\,\dot\tau\Thz\,\mu$ with $\gamma_N\approx 0.7769$. Here, $\xc$ is the critical frequency between DC/BR emission and Compton scattering, with approximations as a function of redshift given in Sect.~3.3.1 of \citep{Chluba2020large}. Every absorption event then removes $\Delta \dot\rho_\gamma/\rho_\gamma\approx \alpha^{-1}_M\,\dot\mu|_{\rm em/abs}$ of energy from the $\mu$-distortion, which is immediately added back to the average temperature, causing a relative temperature shift $\Theta=\Delta T/\Tz$.
Assuming energy conservation, we therefore have the corresponding temperature source term $\dot\Theta|_{\rm em/abs}=-\frac{1}{4\,\gamma_\rho}\,\dot\mu|_{\rm em/abs}\approx \frac{\gamma_N}{4\,\gamma_\rho}\,\xc\,\dot\tau\Thz\,\mu$. Overall this means one has to solve the extended system
\bealf{
\label{eq:simple_ymuT}
\frac{\partial \Theta}{\partial t}
&\approx \gamma_T\,\xc\,\dot\tau\Thz\,\mu,
\qquad
\frac{\partial y}{\partial t}
\approx \frac{1}{4}\dot{\mathcal{Q}}- 4\dot\tau \Thz\,y
\qquad\text{and}\qquad 
\frac{\partial \mu}{\partial t}
\approx\gamma_\rho\,4\dot\tau \Thz\,(4y)-\gamma_N\,\xc\,\dot\tau\Thz\,\mu,
}
with $\gamma_T=\frac{\gamma_N}{4\,\gamma_\rho}\approx 0.1387$, a task that can be easily carried out numerically. Assuming that the photon production process only becomes important when $y$ is already negligible, it is easy to show that $\mathcal{J}_T\approx 1-\mathcal{J}_{\rm bb}$ with distortion visibility $\mathcal{J}_{\rm bb}\approx\expf{-(z/\zmu)^{2.5}}$ and $\zmu=\pot{1.98}{6}$ \citep{Burigana1991,Hu1993}. We then have 
\bealf{
\label{eq:simple_J}
\mathcal{J}_{y}\approx \expf{-4\ysc}, \qquad
\mathcal{J}_{\mu}\approx(1-\mathcal{J}_{y})\,\mathcal{J}_{\rm bb}, \qquad\text{and}\quad
\mathcal{J}_{T}\approx 1-\mathcal{J}_{\rm bb}.
}
As can be seen from Fig.~\ref{fig:Greens_comparison}, these simple approximations already capture the main dependence of the distortion visibility on the injection redshift. The question of the next section is now whether we can improve on this description to also include terms relating to the residual distortion.

\subsection{Preliminaries}
\label{sec:prep-steps}
Neglecting photon production and heating terms, the relevant evolution equation for the distortion, $\Delta n=n-\nbb$, from the blackbody $\nbb=1/(\expf{x}-1)$ in the expanding Universe can be cast into the compact form \citep{Chluba2011therm, Chluba2013Green}
\begin{align}
\label{eq:Main_Eq}
\frac{\partial \Delta n(x)}{\partial \ysc}
&\approx \Theta_{\rm e}\,\Yspec(x)+\KompO\,\Delta n(x)=
\Theta_{\rm e}\,\Yspec(x)+
\DiffO\,\Delta n(x)+\DiffO^* \,A(x)\,\Delta n(x),
\end{align}
where $\Theta_{\rm e}= \frac{\Delta \Te}{\Tz}$ is the relative electron temperature difference and $\KompO=\DiffO+\DiffO^* A$ is the Kompaneets operator, constructed from the diffusion and recoil operators, $\DiffO=x^{-2}\partial_x x^4 \partial_x$ and $\DiffO^*=x^{-2}\partial_x x^4$, with $A=1+2\nbb=(\expf{x}+1)/(\expf{x}-1)$. 
The time variable is the scattering $y$-parameter, $\ysc=\int \dot\tau \Thz \id t$. The problem has been linearised in the distortion, an approximation that will be good unless very large distortions are encountered \citep{Chluba2020large, Acharya2021large}.

For the electron temperature correction, $\Delta \Te=\Te-\Tz$, we assume that Compton equilibrium is reached at all times.\footnote{For the average evolution, this limit is valid on average until very late times corresponding to redshift $z\lesssim 200$.} In the absence of external heating, this means that $\int x^3 \partial_{\ysc} \Delta n\id x\approx 0$, which implies $\Theta_{\rm e}\approx 
\Theta_{\rm eq}$ with \citep[e.g., see][]{Acharya2021FP}
\begin{align}
\label{eq:Compton_eq}
\Theta_{\rm eq}&\approx -\frac{\int x^3\KompO\,\Delta n\id x}{\int x^3 \Yspec(x)\id x}=
\frac{\int (x^4 \partial_x + x^4 A)\,\Delta n\id x}{4\,E_{\nbb}}
\equiv
\frac{\int x^3 w_y\,\Delta n\id x}{4\,E_{\nbb}} 
\end{align}
and the $y$-weight factor $w_y= \Yspec/\Gspec=x A(x)-4=x\frac{\expf{x}+1}{\expf{x}-1}-4$. Since the integrals in Eq.~\eqref{eq:Compton_eq} will appear multiple times, for convenience we introduce
\begin{align}
\label{eq:Compton_eq_average}
\eta_f&
=\frac{\int x^3 w_y(x)\,f(x)\id x}{4\,E_{\nbb}}
\qquad\text{and}\qquad 
\epsilon_f=\frac{1}{\alpha_f}
=\frac{\int x^3 f(x)\id x}{E_{\nbb}}.
\end{align}
The exact integrals that are encountered in our computations can all be given in terms of the Riemann $\zeta$-functions. For the basic spectral shapes we have $\eta_\Gspec=1$, $\eta_{\Yspec}\approx 5.3996$ and $\eta_\Mspec\approx0.4561$, as well as $\epsilon_\Gspec=4$, $\epsilon_{\Yspec}=4$ and $\epsilon_\Mspec=1/1.4007$. For numerical applications we pre-compute all these integrals.

\subsection{Spectral basis and approximate representation of the Kompaneets operator}
\label{sec:SED_basis}
The goal is to find an efficient spectral representation that captures the changes of the spectrum under repeated Compton scattering as described by Eq.~\eqref{eq:Main_Eq}. The simplest decomposition considers the three main spectral types appearing in the thermalisation problem introduced in Eq.~\eqref{eq:basic_spectra}. To build intuition, we discuss this case in some detail, but eventually find it is insufficient. The required refinements are presented right after.

It is instructive to understand the links of these basic spectra to that of the background blackbody spectrum. Both  $\Gspec$ and $\Yspec$ are generated by applications of the boost generator, $\boostO=-x\partial_x$:
\begin{align}
\Gspec(x)
&=\boostO \nbb(x),
\qquad
\Yspec(x)
=\DiffO \nbb(x)
=\boostO(\boostO-3)\nbb(x).
\end{align}
Making the Ansatz $\Delta n=\Theta\,\Gspec(x)+y\,\Yspec(x)+\mu \Mspec(x)$ and inserting back into Eq.~\eqref{eq:Main_Eq} we obtain
\begin{align}
\label{eq:Main_Eq_muy_step_I}
\Theta'\Gspec(x)+y' \Yspec(x)+\mu' \Mspec(x)
&=\Theta_{\rm e}\,\Yspec(y) +\Theta\,\KompO\,\Gspec(x)+y\,\KompO\,\Yspec(x)+\mu\,\KompO\,\Mspec(x),
\end{align}
where the prime indicates the derivative with respect to $\ysc$. The functions
\begin{align}
\label{eq:KG_KM}
K_G\equiv \KompO \Gspec(x)
&=-\Yspec(x) \qquad \text{and} \qquad
K_M\equiv \KompO \Mspec(x)=-\Yspec(x)/\beta_M\equiv -\eta_M \Yspec(x)
\end{align}
nicely map back onto $\Yspec(x)$, while $\KompO\, \Gspec(x)/x=0$ defines the null-space. However, the function $K_\Yspec(x)=\KompO \Yspec(x)$ has contributions that are not spanned by $\Gspec(x), \Yspec(x)$ and $\Mspec(x)$. We can nevertheless enforce a representation of $K_\Yspec(x)$ in terms of $\Gspec(x), \Yspec(x)$ and $\Mspec(x)$. Here, we are mostly interested in intensity functions. Knowing that $K_\Yspec(x)$ does not carry photon number,\footnote{The integral $\int x^2 K_\Yspec(x)\id x =\int x^2 \KompO \Yspec(x) \id x$ vanishes since the Kompaneets operator conserves photon number.} we understand that only $\Yspec(x)$ and $\Mspec(x)$ can contribute in our current basis. 
Since with our finite basis, the representation will not be exact, we can demand that the contribution of $\Mspec$ to follow from energy conservation to improve matters, as discussed below.

We define the matrix elements of the operator $\mathcal{\hat{X}}$ between the two function $F(x)$ and $J(x)$ as\footnote{This is equivalent to taking the integrals over the two intensities $x^3 F$ and $x^3 \mathcal{\hat{X}} J$.}
\begin{align}
\label{eq:scalar_prod}
\langle F | \mathcal{\hat{X}} | J \rangle
\equiv \langle F | \mathcal{\hat{X}} J \rangle
=\int x^3 F(x) \,[x^3 \mathcal{\hat{X}} J(x)] \id x.
\end{align}
To decompose $K_Y=\KompO Y$ in the most simple approach we remap back to the basis using the Ansatz $|\KompO Y\rangle \approx a_0|Y\rangle + a_1|M\rangle$ and solve the system
\begin{subequations}
\label{eq:yM_simplest_bad}
\begin{align}
\langle Y | \KompO | Y \rangle
&\equiv 
\langle Y | K_Y \rangle 
\,\,\approx \langle Y | Y \rangle \,a_0 + \langle Y | M \rangle \,a_1
\\
\langle M | \KompO | Y \rangle
&\equiv
\langle M | K_Y \rangle 
\approx \langle M | Y \rangle \,a_0 + \langle M | M \rangle \,a_1.
\end{align}
\end{subequations}
This system is equivalent to the matrix equation $\vek{b}=M_R\, \vek{a}$, where $b_i=\langle R_i | K_Y \rangle$ for each function of the representation basis, i.e., $R_0=Y$ and $R_1=M$ in the considered case. Similarly, we have the basis mixing matrix $M_{R,{ij}}=\langle R_i | R_j \rangle$ and the corresponding representation coefficients $a_0$ and $a_1$. The solution is then $\vek{a}={M_R}^{-1}\, \vek{b}$, such that $K_Y\approx \vek{R}\cdot\vek{a}$ with $\vek{R}=(\Yspec(x),\Mspec(x))^T$ and $\vek{a}=(a_0,a_1)^T$. Carrying out the projection integrals and inverting the system we obtain $K_Y\approx -8.8169\,\Yspec(x)+40.409\,\Mspec(x)$. However, since we used an incomplete basis, this approximation does not satisfy energy conservation. Carrying out the energy integrals, we find $E_{K_Y}=\int x^3 K_\Yspec(x) \id x \approx -21.598$ by direct integration of the exact function and $E_{K_Y}\approx -8.8169\times 4 + 40.409/1.4007=-6.4178$ from the approximation. Since energy and photon number conservation are the {\it most fundamental} aspects of the thermalisation problem, this is not a solution we can work with.

To fix the problem, we replace the last equation in the system, Eq.~\eqref{eq:yM_simplest_bad}, with the energy conservation equation.
This yields the augmented system
\begin{subequations}
\label{eq:yM_simplest}
\begin{align}
\langle Y | K_Y \rangle 
&\approx \langle Y | Y \rangle \,a_0 + \langle Y | M \rangle \,a_1 
\\
E_{K_Y}
&=E_Y\,a_0+E_M\,a_1,
\end{align}
\end{subequations}
which can still be thought of as $\vek{b}=M_R\, \vek{a}$, but with modified last rows in $\vek{b}$ and $M_R$ according to the energy conservation equation.
By inverting the new system, 
this then yields the improved representation $K_\Yspec(x) \approx -3.4593\,\Yspec(x)-10.871\,\Mspec(x)$. Carrying out the energy integrals, we find $E_{K_Y}\approx -3.4593\times 4-10.871/1.4007=-21.598$, in agreement with the direct integral result.

We have now reformulated the problem once we also determine $\Theta_{\rm e}\approx \Theta_{\rm eq}$. In vector notation, our Ansatz reads $\Delta n=\vek{B}\cdot \vek{y}$, where now we include $\Gspec(x)$ in the basis, i.e., $\vek{B}=(\Gspec(x),\Yspec(x),\Mspec(x))^T$ and $\vek{y}=(\Theta,y,\mu)^T$.
By inserting this Ansatz for $\Delta n$ into Eq.~\eqref{eq:Compton_eq} for the Compton equilibrium temperature perturbation, and carrying out the energy exchange integrals one finds
\begin{align}
\label{eq:Compton_eq_muy_b}
\Theta_{\rm eq}
&\approx \left(\frac{\int x^3 w_y(x)\,\vek{B} \id x}{4 E_{\nbb}}\right) \cdot \vek{y}
=\eta_G\,\Theta+\eta_Y\,y+\eta_M\,\mu\approx
\Theta + 5.3996 y + 0.4561\mu.
\end{align}
Inserting everything back into Eq.~\eqref{eq:Main_Eq_muy_step_I} and collecting terms, with Eq.~\eqref{eq:KG_KM} we then obtain
\begin{align}
\label{eq:Main_Eq_muy}
\Theta'\Gspec(x)+y' \Yspec(x)+\mu'\Mspec(x)
&=\Theta_{\rm e} \Yspec(x) - \Theta\,\Yspec(x)+y\,K_Y-\mu\,\eta_M\Yspec(x)
\nonumber\\&
\approx 1.9403\,y \, \Yspec(x)-10.871\,y \,\Mspec(x).
\end{align}
We note that the terms in the Compton equilibrium temperature $\propto \Theta$ and $\mu$ cancel identically due to the identities in Eq.~\eqref{eq:KG_KM}.
We furthermore comment that Eq.~\eqref{eq:Main_Eq_muy} can be also obtained by directly carrying out the projections onto the basis starting from Eq.~\eqref{eq:Main_Eq}. We show this more formally in Appendix~\ref{app:formal} for the extended basis that is discussed in the next section.

\begin{figure}
	\centering
	\includegraphics[width=0.92\columnwidth]{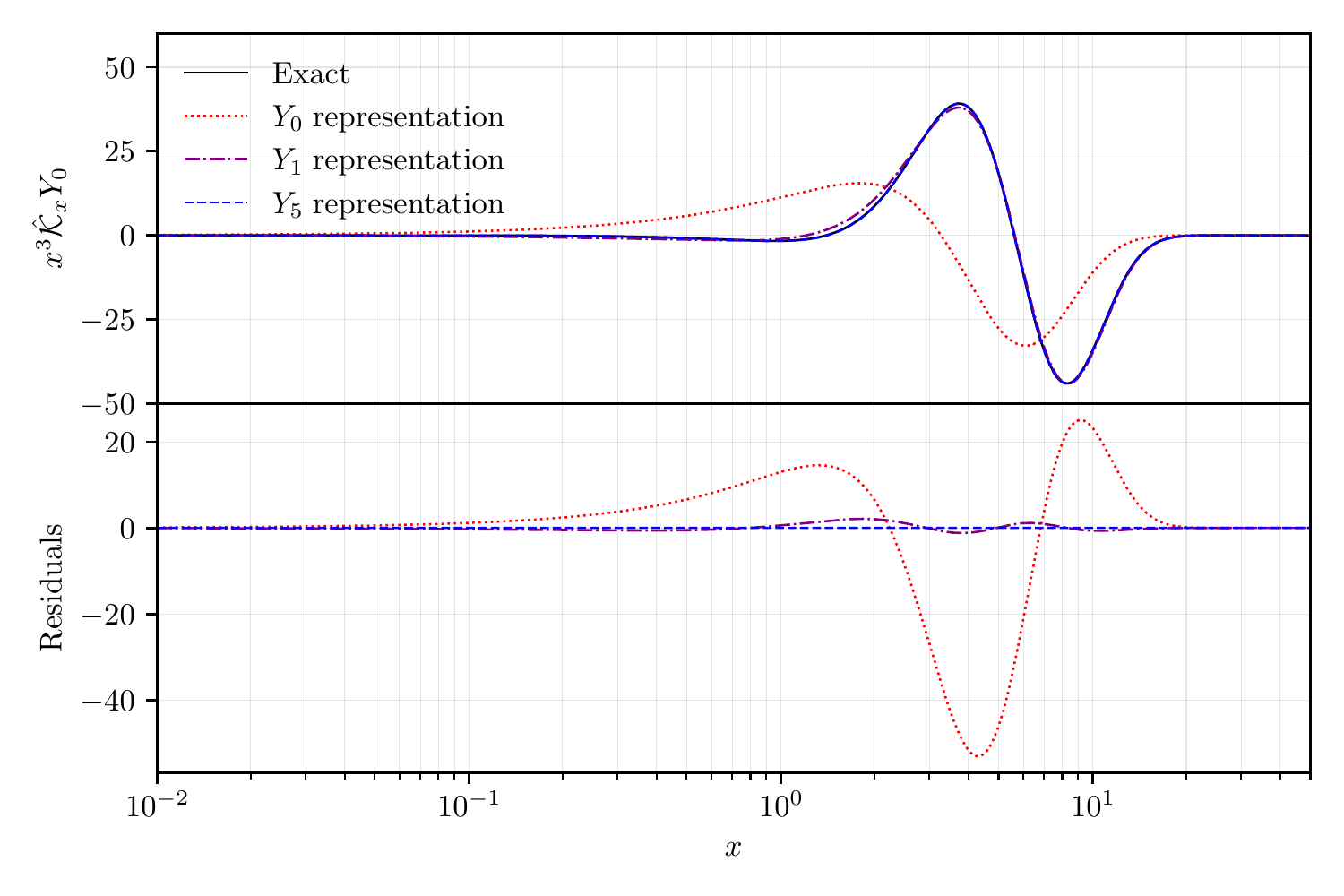}
	\\[-2mm]
	\includegraphics[width=0.92\columnwidth]{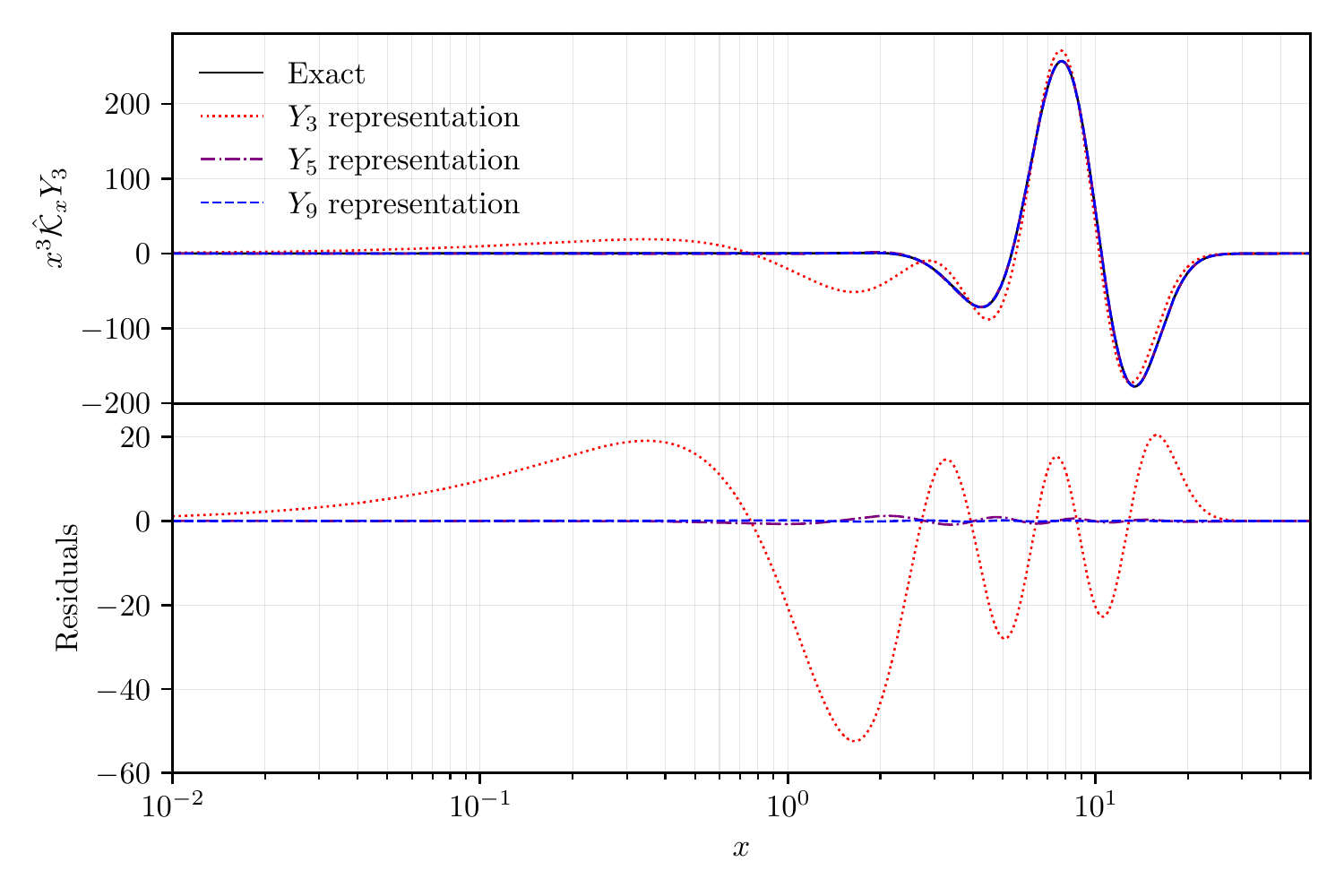}
	\\
	\caption{Distortion shapes $x^3 K_\Yspec(x)$ and $x^3 K_{\Ynspec{3}}(x)$ for various approximations. Here the order refers to the largest term in $Y_k$ that is included [i.e., 0.~order only $\Yspec(x)$ and $\Mspec(x)$; 5.~order includes $M(x)$ and all $Y_k(x)$ up to $Y_5(x)$]. The representations become increasingly accurate the more terms we add to the basis. Typically only poor representation is obtained upon acting on the largest function in the basis -- a problem which is mitigated by the fact that less energy occupy these higher modes in numerical solutions.}
	\label{fig:KY_approx}
\end{figure}

Since the system in Eq.~\eqref{eq:Main_Eq_muy} has to be fulfilled for any $x$ and because the spectral basis is non-degenerate, by comparing coefficients, we obtain the ordinary differential equation (ODE) system
\begin{align}
\label{eq:Main_Eq_muy_finally}
\Theta'&\approx 0, \qquad
y'\approx 1.9403\,y, \qquad 
\mu'\approx -1.9403\, (\epsilon_Y/\epsilon_M)\,y.
\end{align}
with $\epsilon_Y/\epsilon_M\approx 5.6026$. 
While these equations correctly represent the conservation of photon number [only $\Gspec(x)$ carries photon number but $\Theta$ does not change] and also energy (the sum of the energies in $\mu$ and $y$ does not change), they do not yield the correct overall evolution: For the $y$ parameter, the solution is $y(\ysc)\simeq y(0)\,\expf{1.9403 \ysc}$, while we saw in Sect.~\eqref{sec:basic_idea} that it should be more close to $y(\ysc) \approx y(0)\,\expf{-4\ysc}$. {\it What has gone wrong?} The approximate representation of $K_\Yspec(x) \approx -3.4593\,\Yspec(x)-10.871\,\Mspec(x)$ is insufficient, as could have been guessed. This can be appreciated in Fig.~\ref{fig:KY_approx}, where we compare the exact solution of $K_\Yspec(x)$ with various approximations.
In particular the high-frequency part of $K_\Yspec(x)$ is not well-captured by this simplest approximation, a problem that we fix next. 

\subsection{Extension of the basis}
\begin{figure}
	\centering
	\includegraphics[width=0.95\columnwidth]{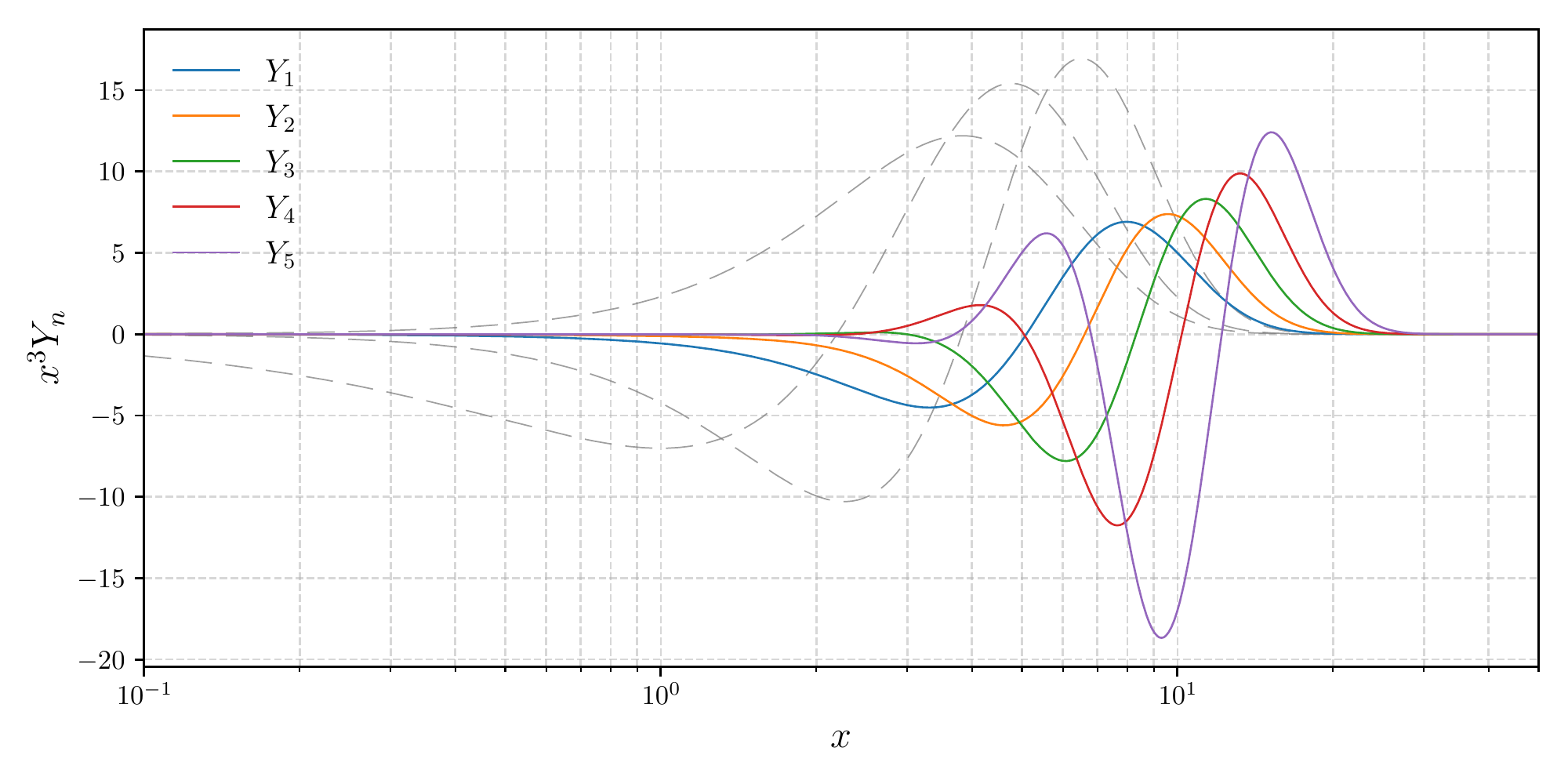}
	\caption{First few basis function $Y_k$ in comparison with the standard distortion shapes (dashed lines, with the positive peak from left to the right relating to $G$, $M$, $Y$, respectively). For increasing $k$ the functions $Y_k$ occupy more and more of the high frequency part of the spectrum.}
	\label{fig:Yn_examples}
\end{figure}
To make progress, we need to extend the spectral basis, $\vek{B}$. One of the natural selections is to use the boost operator $\boostO$ to find the extensions. This is motivated by the fact that $\boostO$ is one of the fundamental operators generating the Kompaneets operator, $\KompO$. It also commutes with the diffusion operator, $[\DiffO,\boostO]=0$, which further supports this choice (see Appendix~\ref{app:operator_props}).
Finally, it appears in log-moment expansions of distortion spectra, which where shown to have useful properties in terms of Gauge-choices \citep{Pitrou2014log, Ota2017}.\footnote{These considerations hold when working with the background spectrum. The picture will be more complicated in the presence of inhomogeneities.}
For $\Gspec(x)$, we have $\boostO \Gspec(x)=\boostO^2 \nbb(x)=
3\Gspec(x)+\Yspec(x)$, which directly maps back onto the old basis. For the boosts of $M(x)$ and $Y(x)$, new spectral shapes are generated. However, since $K_M=\KompO \Mspec(x)=-\eta_M\Yspec(x)$ already maps back onto our basis, for now we only need to think about extensions based on the functions 
\begin{align}
\Ynspec{k}(x)&=(1/4)^k\boostO^k \Yspec(x).
\end{align}
These functions can be readily computed using {\tt Mathematica} or through the combinatoric sums given in Appendix~\ref{app:operator_props}, and are illustrated for a few cases in Fig.~\ref{fig:Yn_examples}. 
The $Y_k$ are similar to those functions appearing in asymptotic expansions of the SZ effect \citep{Itoh98, Sazonov1998, Challinor1998, Chluba2012SZpack} and can furthermore be found in perturbative expansions of the photon transfer problem \citep{Pitrou2009, Khatri2012mix, Ota2014, Haga2018}.
Note that $\Ynspec{0}(x)\equiv \Yspec(x)$. We also added the factor of $(1/4)^k$ to make each of the $Y_k$ more comparable in amplitude. This choice also ensures $\epsilon_{Y_k}\equiv 4$.
These functions all conserve photon number ($\int x^2 \Ynspec{k}(x)\id x=0$) and hence provide a natural extension of the simple $Y$ and $M$ basis. As we will see in paper II, these also naturally appear once Doppler-driving in the perturbed Universe is included. 

\subsection{Generalization of the ODE system}
\label{sec:general_ODE}
In this section, we outline the basic approach for obtaining a generalized ODE system in the extended basis. By deciding about how many $Y_k$ we include in the Ansatz for $\Delta n$, we have to determine the representations for each of the\footnote{These arise from applying $\KompO$ to the Ansatz for $\Delta n$.} $K_{Y_k}=\KompO\,Y_k$ within this basis. To simplify the notation, let us again write the extended representation basis as a vector $\vek{R}(x)=(\Yspec(x), \Ynspec{1}(x), \ldots, \Ynspec{N}(x),\Mspec(x))^{T}$. We will denote $R_0=\Ynspec{0}\equiv Y$ and $R_{N+1}=M$, with all the other $R_k=Y_k$ in between. We then make the Ansatz $K_{Y_k}\approx \vek{R}\cdot \vek{a}_{Y_k}$ with $\vek{a}_{Y_k}$ denoting the coefficients of each term in the representation basis, $\vek{a}=(a_0, a_1, \ldots,a_N,a_{N+1})^T$. As above, we now have to compute the projection of $K_{Y_k}$ onto each of the $R_i$. To ensure energy conservation, we will again determine $\mu$ using the energy integrals.\footnote{We could really replace any one equation using energy conservation.} This then yields the following system of equations that determines the representation vector $\vek{a}_{Y_k}$:
\begin{align}
\langle Y | K_{Y_k} \rangle
&\approx \langle Y | Y \rangle \,a_{Y_k,0}
+\langle Y | \Ynspec{1} \rangle \,a_{Y_k,1}
+\ldots 
+ \langle Y | Y_{N-1} \rangle \,a_{Y_k,N-1}
+\langle Y | Y_{N} \rangle \,a_{Y_k,N}
+\langle Y | M \rangle \,a_{Y_k,N+1}
\nonumber\\
\langle \Ynspec{1} | K_{Y_k} \rangle
&\approx \langle \Ynspec{1} | Y \rangle \,a_{Y_k,0}
+\langle \Ynspec{1} | \Ynspec{1} \rangle \,a_{Y_k,1}
+\ldots 
+\langle \Ynspec{1} | Y_{N-1} \rangle \,a_{Y_k,N-1}
+\langle \Ynspec{1} | Y_{N} \rangle \,a_{Y_k,N}
+\langle \Ynspec{1} | M \rangle \,a_{Y_k,N+1}
\nonumber\\[-1mm]
\vdots\,\quad\quad &\approx
\quad \quad  \vdots
\\ \nonumber
\langle Y_{N} | K_{Y_k} \rangle
&\approx\langle Y_{N} | Y \rangle \,a_{Y_k,0}
+\langle Y_{N} | \Ynspec{1} \rangle \,a_{Y_k,1}
+\ldots 
+\langle Y_{N} | Y_{N-1} \rangle \,a_{Y_k,N-1}
+\langle Y_{N} | Y_{N} \rangle \,a_{Y_k,N}
+\langle Y_{N} | M \rangle \,a_{Y_k,N+1}
\\ \nonumber
E_{Y_k}&\approx
E_Y\,a_{Y_k,0}+E_{\Ynspec{1}}\,a_{Y_k,1}+\ldots+ E_{Y_{N-1}}\,a_{Y_k,N-1}+ E_{\Ynspec{N}}\,a_{Y_k,N}+E_M\,a_{Y_k,N+1},
\end{align}
The last equation is the energy conservation equation to determine the coefficient of $\Mspec(x)$. We thus have a matrix equation of the form $\vek{b}_{K_{Y_k}}=M_R \, \vek{a}_{K_{Y_k}}$, which we can solve for $\vek{a}_{K_{Y_k}}$ given a finite representation basis. We again highlight the fact that the system was obtained using the energy conservation equation. The matrix $M_R$ is therefore again nearly equivalent to the full basis mixing matrix $M_{R,ij}=\langle R_i | R_j \rangle$. However, the last equation is replaced by the energy conservation equation, even if not explicitly distinguished in the notation.

As an example, if we choose $\vek{R}(x)=(\Yspec(x), \Ynspec{1}(x),\Mspec(x))^{T}$, we only have to determine the representations for $\KompO Y$ and $\KompO \Ynspec{1}$. Solving the corresponding systems of equations then yields
\begin{subequations}
\begin{align}
\label{eq:rep_yy1mu}
K_\Yspec(x)&\approx 2.4717\,\Yspec(x)-8.4907\,\Ynspec{1}(x)+3.4698\,\Mspec(x)
\equiv\vek{R}\cdot \vek{a}_{Y}
\\
K_{\Ynspec{1}}(x)&\approx 
28.134\,\Yspec(x)-26.125\,\Ynspec{1}(x)-55.089\,\Mspec(x)\equiv \vek{R}\cdot \vek{a}_{\Ynspec{1}}.
\end{align}
\end{subequations}
Looking at Fig.~\ref{fig:KY_approx}, we see that now the match with the exact result for $K_\Yspec(x)$ is already very good. 
We note that adding more terms to the basis evidently changes {\it all} the coefficients of the solution, and also improves the result for the correspondence (see Fig.~\ref{fig:KY_approx}). 
Inserting this back into Eq.~\eqref{eq:Main_Eq} and using $\eta_{\Ynspec{1}}\approx 7.8246$ in Eq.~\eqref{eq:Compton_eq}, after collecting coefficients we then find
\begin{align}
\label{eq:Main_Eq_muyy1}
&\Theta'\Gspec(x)+y' \Yspec(x)+y_1' \Ynspec{1}(x)+\mu'\Mspec(x)
=\Theta_{\rm e}\,\Yspec(x)- \Theta\,\Yspec(x)+y\,K_\Yspec(x)+y_1\,K_{\Ynspec{1}}(x)-\mu\,\eta_M \Yspec(x)
\\[1mm] \nonumber
&\qquad\approx 
(7.8714 y +35.958 y_1) \, \Yspec(x)
-(8.4907 y + 26.125 y_1) \,\Ynspec{1}(x) 
+(3.4698 y -55.089 y_1) \,\Mspec(x).
\end{align}
In Appendix~\ref{app:formal} we give an alternative derivation that avoids the intermediate step of first representing the $K_{Y_k}$ in terms of the basis. However, mathematically this is equivalent.
By comparing the coefficients, one can again obtain a system for the evolution of $\Theta, y, y_1$ and $\mu$.
The solution of this system now has the correct main properties. It conserves photon number and energy and leads to a solution\footnote{This can be seen when assuming that the coefficient of $\Ynspec{1}$ evolves under quasi-stationary conditions. This implies the condition $8.4907 y + 26.125 y_1\approx 0$ resulting in $y_1^{\rm qs}\approx -0.32550 \, y$, which yields the desired result similar to Eq.~\eqref{eq:simple_ymu}.} $y(\ysc)\simeq y(0)\,\expf{-3.8\,\ysc}$.
Indeed this is very close to the correct Green's function solution that neglects any residual distortion contributions. However, the precision can be improved by further extending the spectral basis (see Fig.~\ref{fig:KY_approx}).

Below we will give the solutions for systems that include up to $\Ynspec{15}(x)$ in the basis. This already provides a very accurate approximation for the exact Green's function. The related system can be readily generated using {\tt Mathematica} following the procedure above. Schematically, we can then express the effect of the Kompaneets operator on the distortion in the form 
\begin{align}
\label{eq:Main_Eq_muyy1_scheme}
\Delta n'=\Theta_{\rm e}\,Y+\KompO\,\Delta n
\qquad \longleftrightarrow \qquad
\vek{y}'&\approx M_{\rm K}\,\vek{y}
\end{align}
with $\vek{y}=(\Theta, y, y_1, \ldots, y_N,\mu)^{T}$ and where $M_{\rm K}$ is the Kompaneets mixing matrix that directly depends on the chosen spectral basis.\footnote{We will provide the system for up to $y_{15}$ under \url{www.Chluba.de/CosmoTherm}.}
Even order systems are omitted, as they are found to be numerically unstable.\footnote{One can change the weight function in the definition of the scalar-product, Eq.~\eqref{eq:scalar_prod}, to remedy this issue, but we did not explore this option any further.}
We suspect this is due to the second order nature of the Kompaneets operator, but have no additional prove for this.
The solution at any moment is then $\Delta n(x, \ysc)\approx \vek{B}(x)\cdot\vek{y}(\ysc)$ with the full spectral basis $\vek{B}(x)=(\Gspec(x), \Yspec(x), \Ynspec{1}(x), \ldots, \Ynspec{N}(x),\Mspec(x))^{T}$.

\subsection{Adding the effect of photon production and heating}
To add the effect of photon production by double Compton (DC) and Bremsstrahlung (BR), we make use of the fact that once these become important, the $y_k$ will be extremely short-lived (i.e., decay quickly, $y_k\rightarrow 0$). In this case, we can neglect the role of the $Y_k$'s for photon production and the analytic results for the $\mu$-distortion evolution can be used \citep{Sunyaev1970mu, Chluba2014}. 
The net photon emission and absorption term has the explicit form \citep{Hu1993, Chluba2011therm, Chluba2014}
\begin{align}
\frac{1}{\dot{\tau}}\,\frac{\partial n_0}{\partial t}\Bigg|_{\rm em/abs}&=\frac{\Lambda(x, \The,\Thz)\,\expf{-x\,\Thz/\The}}{x^3}\left[1-n_0\left(\expf{x\,\Thz/\The}-1\right)\right]
\nonumber
\approx - \frac{\Lambda(x, \Thz)\,(1-\expf{-x})}{x^3}\Delta n_0+
\frac{\Lambda(x, \Thz)}{x^2}\,\nbb\,\Theta_{\rm e}.
\end{align}
In the last step, we again linearised the problem with respect to the distortion [and $\Theta_{\rm e}\simeq \mathcal{O}(\Delta n)$]. 
The DC and BR emissivities can be computed accurately using {\tt DCpack} \citep{Ravenni2020DC} and {\tt BRpack} \citep{BRpack}.

As already explained in Sect.~\ref{sec:basic_idea}, we can think of the effect that photon emission and absorption has on the distortion as a {\it redistribution} between $\mu$ and $\Theta$. Overall this means
\begin{align}
\label{eq:zeroth_equation_source}
\frac{\partial n_0}{\partial \ysc}\Bigg|_{\rm em/abs}&
\longleftrightarrow \gamma_T\,\xc\,\mu\,\Gspec(x)-\gamma_N\,\xc\,\mu\,\Mspec(x),
\end{align}
as in Eq.~\eqref{eq:simple_ymuT}. This greatly simplifies the thermalisation problem, essentially converting the collision term into a source-sink term with built-in energy conservation.

To also add the effect of external heating, we assume that the distortions are generated through a $y$-distortion source, $y'=(1/4)\,\mathcal{Q}'$, where $\mathcal{Q}'=\id \mathcal{Q}/\id \ysc=(\dot{\tau}\Thz)^{-1}\id \mathcal{Q}/\id t$ in this context. For energy release scenarios, this will be a very good approximation in the pre-recombination era, since heat that is transferred to the baryons quickly reaches the photons through Compton scattering \citep[e.g., see][]{Sazonov2001, Chluba2011therm}. 
The factor of $\alpha_Y=1/4$ converts the change of the relative energy density into the $y$-parameter. Together we then have 
\begin{align}
\label{eq:Main_Eq_photon_injection}
&\Delta n'=\Theta_{\rm e}\,Y+\KompO\,\Delta n
+\Delta n'|_{\rm em/abs}+\Delta n'|_{\rm h}
\quad \longleftrightarrow \quad
\vek{y}'\approx M_{\rm K}\,\vek{y}
+\vek{D}+\frac{\vek{Q}'}{4},
\nonumber\\
&\qquad\vek{D}
=
\left(
     \gamma_T\xc\,\mu, 0, 0, \ldots, 0, -\gamma_N\xc\,\mu
\right)^T,
\quad
\vek{Q}'
=\left(0, \mathcal{Q}', 0, \ldots, 0, 0\right)^T.
\end{align}
This equation now allows us to account for the effects of external heating and emission/absorption with the source vectors, $\vek{Q}'$ and $\vek{D}$, respectively.
Refinements to the treatment of photon emission and absorption that include the effects of $Y_k(x)$ as well as other corrections to the leading order terms can in principle be added following the method of \cite{Chluba2014}; however, for now we stop with this simple description, emphasizing again that most of the thermalisation of distortions occurs deep into the $\mu$-era, when these effects are expected to be small.

\subsection{Solutions for the Green's function after single injection}
\begin{figure}
\centering
\includegraphics[width=\columnwidth]{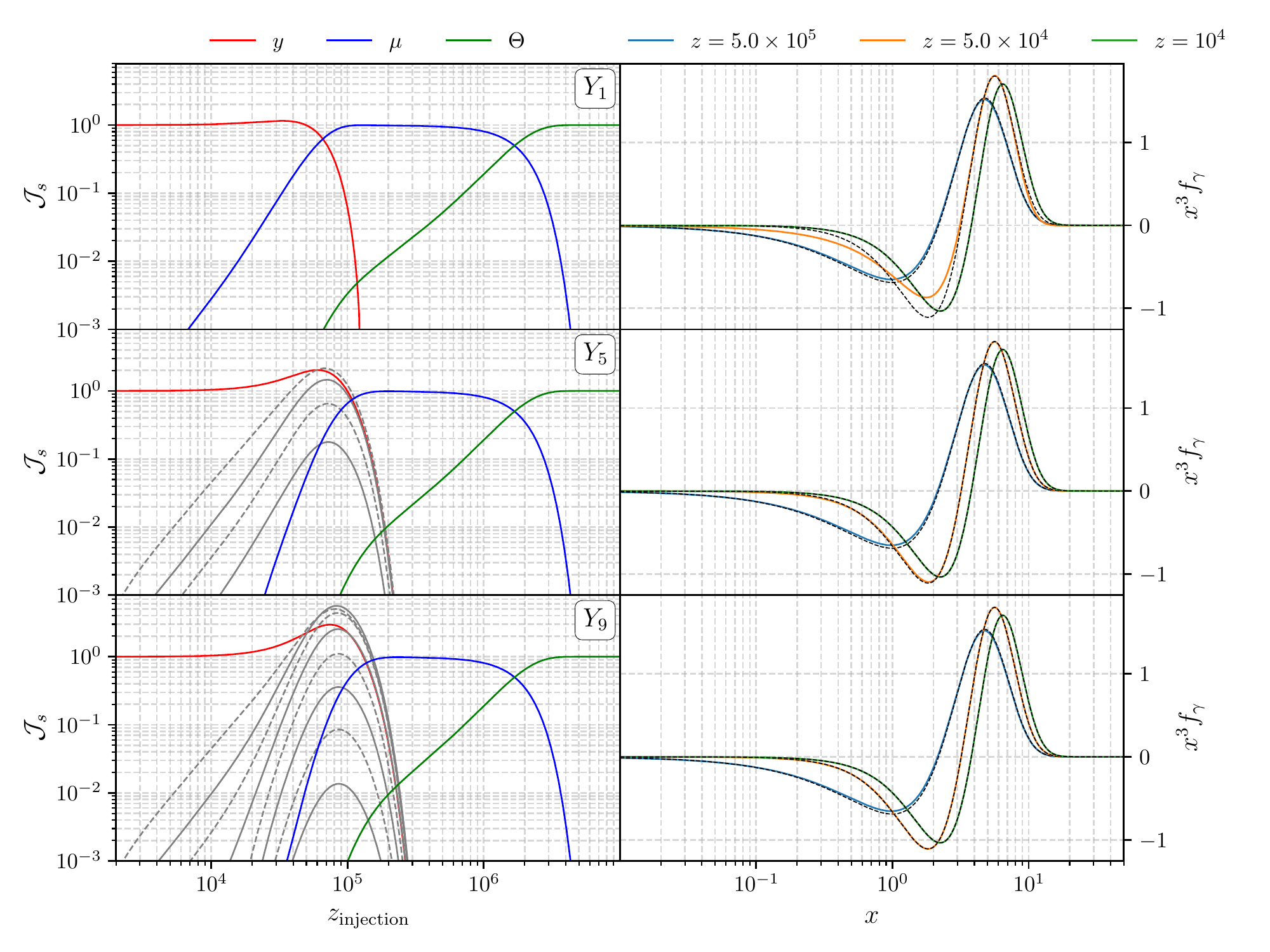}
\caption{A figure showing the iterative improvements of augmenting the $\Ynspec{k}$ basis. The rows show the branching ratios across redshifts (left) and final spectrum at various energy injection redshifts (right) for $N_{\rm max}=1,\,5,\,9$. Gray lines in the branching ratio plots correspond to the $\Ynspec{k>0}$ coefficients, and dotted lines are negative values. Dotted black lines in the spectrum plots show the full results performed with {\tt CosmoTherm}.}
\label{fig:branch_greens_YNmax}
\end{figure}
\begin{figure}
\centering
\includegraphics[width=0.98\columnwidth]{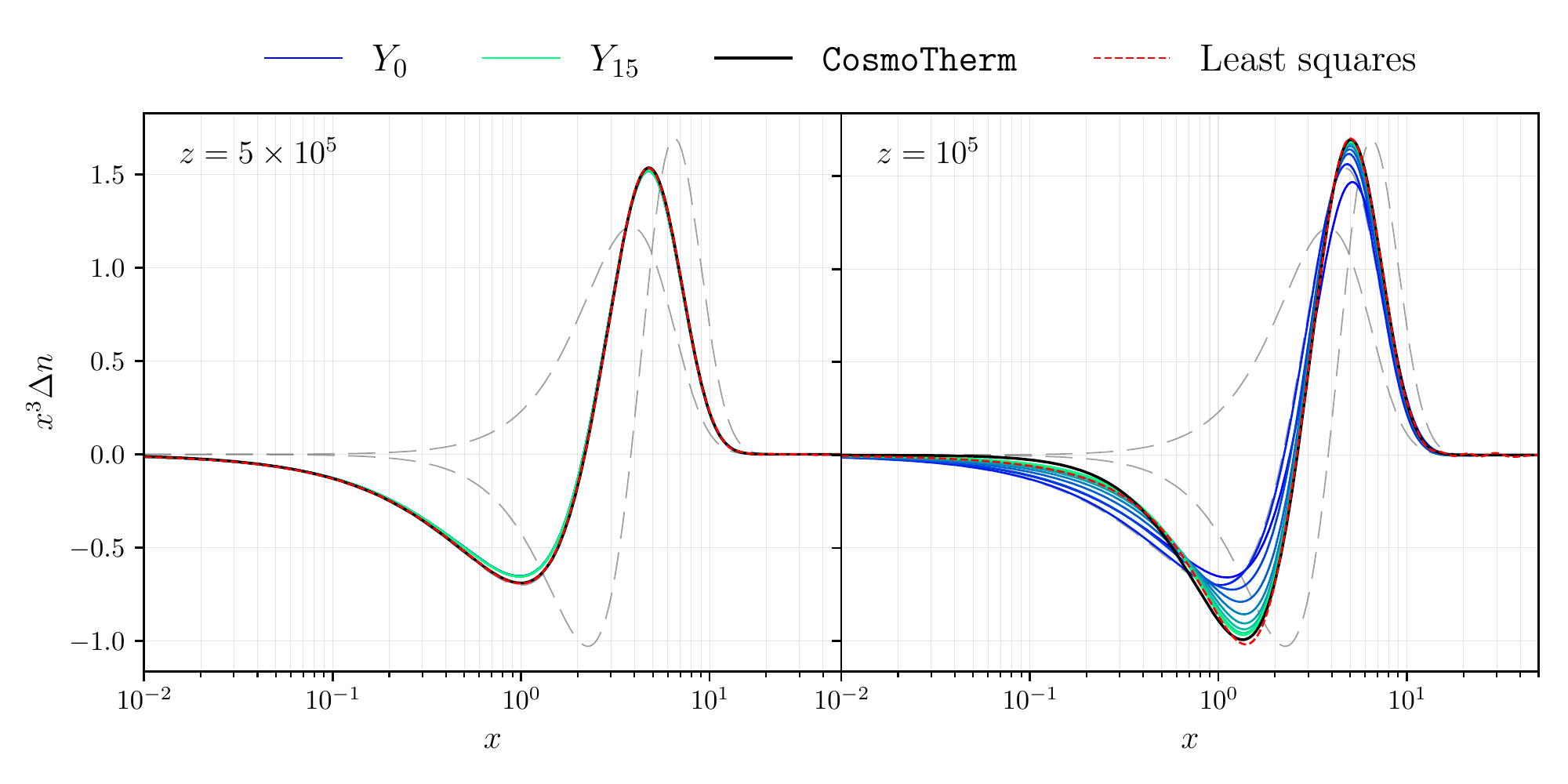}
\\
\caption{A figure showing two characteristic time slices where the full solution is poorly approximated. At $z=5\times 10^5$ (left) the approximate basis still has an excessive contribution from $\Gspec$. At $z=10^5$ (right) the different order of bases approach the residual distortion shape, but perform especially badly at low-frequencies. Also shown is a least-squares fit (red dotted line) using the $Y_{11}$ basis.}
\vspace{-3mm}
\label{fig:YN_bad_convergence}
\end{figure}
The thermalisation Green's function has been successfully used to represent the spectral distortion shapes from continuous heating \citep{Chluba2013Green, Chluba2013fore, Chluba2013PCA}. With the above description we can reproduce the Green's function to high precision, as we show now.
For this, we model the scenario of single injection in Eq.~\eqref{eq:Main_Eq_photon_injection} and introduce a narrow Gaussian heating rate at $z_{\rm injection}$ (or alternatively set an effective initial condition for $y$ at that redshift). Allowing this to evolve under successive scatterings we study the state of the system $\vek{y}(\zf)$ at the final redshift $\zf$, and extract the corresponding spectral shape.

The result of this calculation for various basis sizes is shown in Fig.~\ref{fig:branch_greens_YNmax}. Also shown is a comparison to the exact result of {\tt CosmoTherm}, which performs an analogous calculation by directly binning the frequency space. The latter approach can be thought of as applying a "top hat" basis in $x$ to the same formalism discussed in Sect.~\ref{sec:ODE_Greens}, and thus is more precise at the cost of tracking thousands of equations simultaneously. Despite the relative simplicity of the treatment derived here it is possible to capture the transition from temperature shift to $y$-distortion through the intermediate $\mu$ and residual eras accurately. The residual era in particular is captured by the expanded basis $Y_k$, with $N_{\rm max}=9$ already yielding very accurate results.

It is noteworthy that with the inclusion of the new spectral shapes $Y_k$ the definitions of $y$ and $\mu$ have a degree of degeneracy. This is most notable in the recession of the $\mu$-era with increasing $N_{\rm max}$ and the existence of $\mathcal{J}_y>1$ in the residual era (energy conservation is ensured by cancellation with the negative $\mathcal{J}_{y_k}$). This apparent \textit{arbitrary} labelling of energy with different coefficients is not problematic, since the real physical observable that must converge is the spectrum, which indeed remains stable as seen in the right panels of Fig.~\ref{fig:branch_greens_YNmax}. This physical observable will itself be projected onto some beneficial spectral shapes, as discussed in Sect.~\ref{sec:observables}, which depend on the characteristics of observing instrument \cite{Chluba2013PCA, Lucca2020} or other theoretical choices.

While the spectra in Fig.~\ref{fig:branch_greens_YNmax} show each snapshot being captured accurately, we note that the precise timings of the transition from one phase to the next appear slightly delayed relative to the full {\tt CosmoTherm} calculation.\footnote{Videos illustrating the solution will become available at \url{www.Chluba.de/CosmoTherm}.} In Fig.~\ref{fig:YN_bad_convergence} we show two time slices in the transition phases $T\rightarrow\mu$ and $\mu\rightarrow y$, again with their respective {\tt CosmoTherm} comparison and an optimised least squares fit to the full solution using the approximate basis of spectral functions. At $z=5\times 10^5$ (left panel) we see the approximate solution having part of a temperature shift while the full solution is almost a pure $\mu$ distortion. Seeing that the optimised fit reproduces the {\tt CosmoTherm} solution well, we conclude that the approximate treatment slightly overestimates the thermalisation timescale. As explained in \cite{Chluba2014}, several additional aspects that are not captured by the simple treatment here do matter at the level of a few percent. By more carefully treating the DC and BR thermalisation rate, which will lead to a refined scaling of $\xc$ with time, one can probably improve the treatment; however, for our purpose the current approximation shall suffice, and refinements are left to future work.

In the right panel of Fig.~\ref{fig:YN_bad_convergence}, we also see a snapshot at $z=10^5$. It is apparent that the approximate basis approaches the full solution, but does not capture it fully. 
For comparison, an optimised fit (which allows one to smooth over any time-dependent mismatch) is shown but also fails to exactly reproduce the curve in this case. We can therefore conclude the basis only has enough freedom to capture some --- but not all --- of the nuances of the residual era.
Departures from the solution are visible at low and intermediate frequencies (i.e., $x\lesssim 2$) owing to the nature of the chosen basis (see Fig.~\ref{fig:Yn_examples}) and our focus on energy conservation, which is driven by the high-frequency tail. 
Additional work on the optimal basis will likely remedy these limitations; however, we highlight that our treatment already greatly improves the modeling of the residual era, which is barely captured using a simple $y$ and $\mu$ approximation. Hence, we again shall be content with the performance of the ODE treatment and focus on applications to anisotropic distortions as the main next step (paper II).

\section{Defining spectral distortion observables}
\label{sec:observables}
In Fig.~\ref{fig:branch_greens_YNmax} we saw that the amplitude of the $y$ distortion changes depending on the other amplitudes within the expanded basis while leaving the actual photon spectrum unchanged. In this section, we will formalize and further discuss this phenomenon in the context of changing basis. Heuristically, we can see the space of valid spectra as an abstract vector space, and as such choose a basis for this space. Provided the spectrum is a continuous function we expect a formal basis to be infinite, but computationally a finite basis can suffice, if chosen well. The bottom line statement we emphasise and highlight here then is that the underlying physics will be (and must be) independent of the choice of basis, where the physics here is captured only by the full photon spectrum and not any individual branching ratio or transfer function.

This gives the freedom to choose a basis which suits a given purpose most appropriately. In this light we will introduce two new bases, which are useful for packaging and exporting the results of the $Y_k$ basis or {\it computation basis}.
At a given spectral sensitivity, only a finite number of spectral parameters will be directly measurable, and it is moot to attempt determining the amplitudes (or power and cross-power spectra) for all the spectral parameters inherent to the computation basis. The other two bases introduced here are guided by the principle to compress the information in the spectrum and prepare for easily extracting and interpreting the physics in observations.

Although from our discussion it is clear that a better computation basis which captures all the spectral complexity at low frequencies may exist, we are now interested in finding alternative representations for the space spanned by our $Y_k$ basis. 
As explained in \cite{Chluba2013PCA}, for a given experimental setting (e.g., frequency coverage and channel sensitivities) one can ask which spectral shapes are best constrained aside from the standard distortions. These spectral shapes can be determined using a principal component analysis. Mathematically, this can be thought of as an expansion of the spectrum into $\mu$, $y$ and $\Theta$ plus some additional spectral parameters, $r_i$, to describe the residual distortion shapes.\footnote{In \cite{Chluba2013PCA} the residual distortion amplitudes are referred to as $\mu_i$, but we shall use a new nomenclature henceforth.} The residual distortion shapes are the principal spectral components spanning the residual distortion space and can be ranked by their observability, defining the {\it observation basis}. 
Denoting the residual distortion eigenspectra as $\vek{S}^{(k)}$, we find
\begin{align}
\Delta I_i&=\int B_i(\nu)\,\Delta I_\nu \id \nu=
\Theta \,\Delta I^G_i+y \,\Delta I^Y_i+\mu\,\Delta I^M_i + \Delta R_i,
\qquad\text{and}\qquad
\Delta R_i=\sum_{k=1} r_k\, S_i^{(k)}
\end{align}
where $\Delta I_\nu=2h\nu^3/c^2 \Delta n_\nu$ is the intensity corresponding to $\Delta n_\nu$, which is integrated over the bandpass, $B_i(\nu)$. In our computation we shall use a simple top-hat bandpass centered around frequency $\nu_i$ with a width $\Delta \nu_i$. Similarly, $\Delta I^G_i$, $\Delta I^Y_i$ and $\Delta I^\mu_i$ are the band-averaged versions of the corresponding $G$, $Y$ and $M$ intensities.
The (band-averaged) residual distortion, $\Delta R_i$, space is orthogonal to $M$, $Y$ and $G$, for the selected instrumental configuration. 
Since the binned spectral shapes can all be thought of as simple vectors, we can directly obtain the $\mu$, $y$, $\Theta$ and $r_k$ values for any distortion signal as 
\begin{align}
\begin{pmatrix}
\Theta_{\rm o}
\\[-0.5mm]
y_{\rm o}
\\[-0.5mm]
\mu_{\rm o}
\end{pmatrix}
&=
\begin{pmatrix}
\Delta \vek{I}^G\cdot \Delta \vek{I}^G\;\;
&
\Delta \vek{I}^G\cdot \Delta \vek{I}^Y\;\;
&
\Delta \vek{I}^G\cdot \Delta \vek{I}^M
\\[-0.5mm]
\Delta \vek{I}^Y\cdot \Delta \vek{I}^G\;\;
&
\Delta \vek{I}^Y\cdot \Delta \vek{I}^Y\;\;
&
\Delta \vek{I}^Y\cdot \Delta \vek{I}^M
\\[-0.5mm]
\Delta \vek{I}^M\cdot \Delta \vek{I}^G\;\;
&
\Delta \vek{I}^M\cdot \Delta \vek{I}^Y\;\;
&
\Delta \vek{I}^M\cdot \Delta \vek{I}^M
\end{pmatrix}^{-1}
\,
\begin{pmatrix}
\Delta \vek{I}^G\cdot \Delta \vek{I}
\\[-0.5mm]
\Delta \vek{I}^Y\cdot \Delta \vek{I}
\\[-0.5mm]
\Delta \vek{I}^M\cdot \Delta \vek{I}
\end{pmatrix}
\qquad\text{and}\qquad
r_k=\frac{\vek{S}^{(k)}\cdot \Delta \vek{I}}{\vek{S}^{(k)}\cdot\vek{S}^{(k)}}.
\end{align}
This assumes that the covariance of the spectral bands is diagonal, but extensions can be readily given. The residual distortion parameters, by construction, will {\it only} receive contributions from the $y_i$ of our computation basis, while $\Theta_{\rm o}$, $\mu_{\rm o}$ and $y_{\rm o}$ will be a superposition of the $\Theta$, $y$ and $\mu$ values in the previous basis with extra contributions from the $y_i$.
The relevant {\it rotation} of the basis can be precomputed (see Sect.~\ref{sec:basis_change}).
Given the observation basis $\vek{S}^{(k)}$ we can therefore usually compress the information into fewer observational parameters, as we show below. 

\begin{figure}
\centering
\includegraphics[width=0.85\columnwidth]{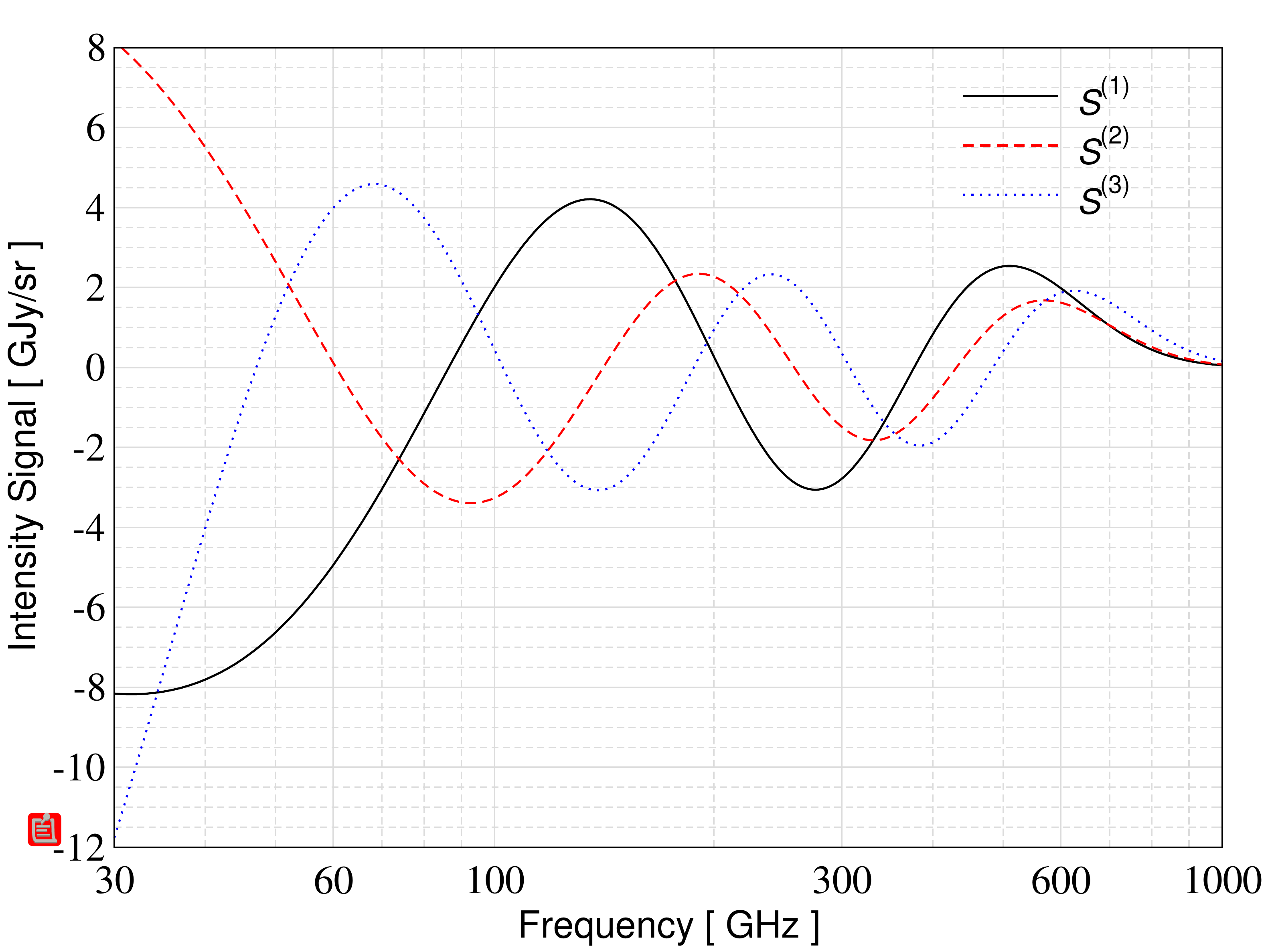}
\caption{First three residual distortion eigenmodes obtained for $\Delta \nu=1\,{\rm GHz}$ in the range $\nu_{\rm min}=30\,{\rm GHz}$ to $\nu_{\rm max}=1000\,{\rm GHz}$. These signals are orthogonal to the standard $G$, $Y$ and $M$ spectra and also among each other. They have all been normalized to carry an energy of $\Delta \rho_\gamma/\rho_\gamma=4$.}
\label{fig:PCAs}
\end{figure}

In Fig.~\ref{fig:PCAs}, we show the first few $\vek{S}^{(k)}$ used in our computations below. The basis was created assuming constant channel sensitivity and channel widths $\Delta \nu=1\,{\rm GHz}$ in the range $\nu_{\rm min}=30\,{\rm GHz}$ to $\nu_{\rm max}=1000\,{\rm GHz}$, mainly for illustration. We normalized all of these to carry $\Delta \rho_\gamma/\rho_\gamma = 4$ of energy. This choice makes them comparable in amplitude to the standard distortion shapes and the level of the corresponding residual distortion parameter gives away its relative importance.
Creating the optimal distortion eigenmodes for more realistic experimental configurations is straightforward following the procedure outlined in \cite{Chluba2013PCA, Lucca2020}. We can see that the distortion eigenmodes exhibit an increasing number of nodes, reminiscent of other orthogonal functions sets. In applications, this will typically lead to the corresponding residual distortion parameter, $r_i$, decreasing in amplitude.

\begin{figure}
\centering
\includegraphics[width=\columnwidth]{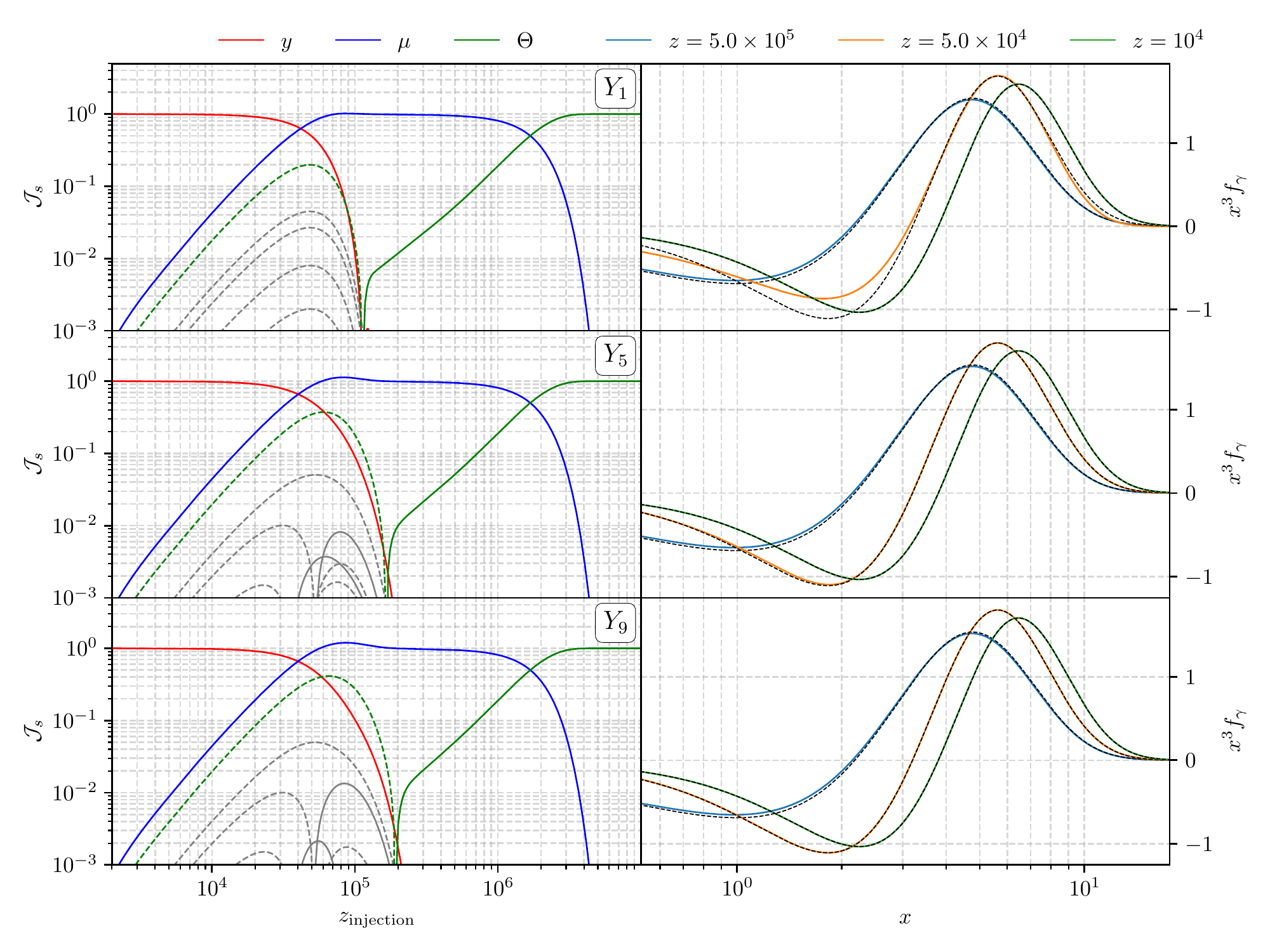}
\\
\caption{As for Fig.~\ref{fig:branch_greens_YNmax}, but now with results cast into the ``observation'' basis. Notice how now the $y$ and $\mu$ amplitudes are stable with increasing basis size. The spectra cover a smaller frequency range, as dictated by realistic observational scenarios, however they do not otherwise change compared to the computational basis. The residual era shows an effective negative temperature shift to achieve the correct spectral shape.}
\label{fig:fig:branch_greens_obs}
\end{figure}

In Fig.~\ref{fig:fig:branch_greens_obs} we illustrate how this mapping to the observation basis modifies the appearance of the branching ratios (left panels) and photon spectra (right panels). We immediately see that the $y$ distortion does not take on a relative energy contribution $>1$ as it did in the computation basis (Fig.~\ref{fig:branch_greens_YNmax}). Further to this point, the total amplitude of $y$ and $\mu$ are more stable for increasing $y_{n>0}$, revealing that the modelling of spectral evolution is improving with basis size, but without the usual incurred \textit{coefficient ambiguity} as a trade-off. The spectra cover a smaller frequency range, as discussed above, but otherwise show no significant departure from the result of the computational basis. Recall the statement that the bottom line physical results -- the spectrum -- are independent of chosen basis.

\subsection{Efficient change of the basis}
\label{sec:basis_change}
To accelerate the calculation we can precompute all `rotations' from one basis to the other given the distortion vectors (which depend on the experimental setting). Algorithmically, we have to bin all the involved spectra from the various bases and then compute the relevant mixing matrices and subsequently invert the problem. This then defines the mixing matrix $L$, which maps $\vek{y}=(\Theta, y, y_1, ..., y_N, \mu)$ to $\vek{o}=(\Theta_{\rm o}, y_{\rm o}, r_1, ..., r_M, \mu_{\rm o})$ as $\vek{o}=L\,\vek{y}$. The dimension of the two spaces need not be the same, with the observation basis having a lower dimension given that the observability of various independent signal modes is usually reduced.

\begin{figure}
\centering
\includegraphics[width=0.75\columnwidth]{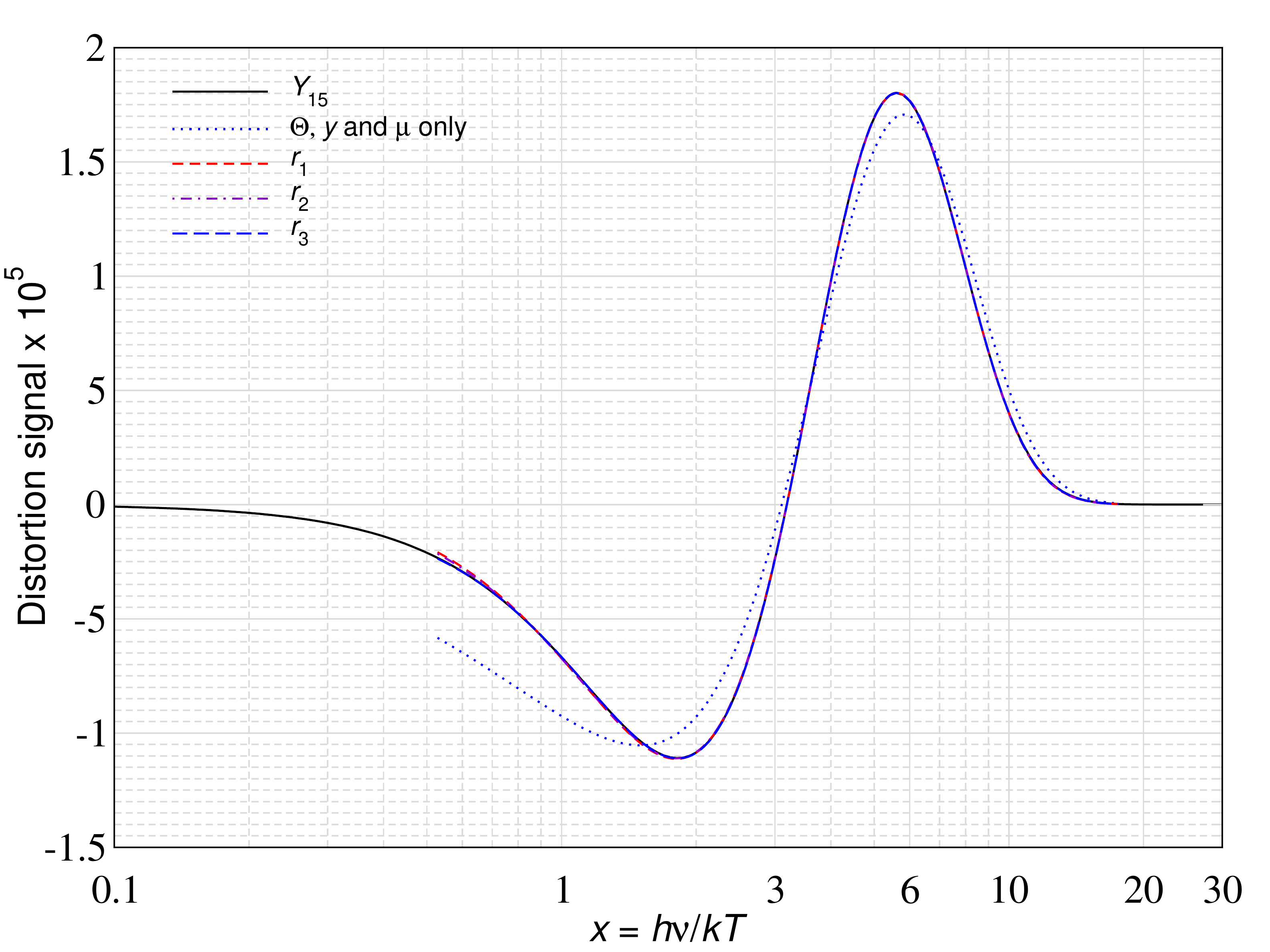}
\\
\includegraphics[width=0.75\columnwidth, trim={0cm 9cm 0 0cm},clip]{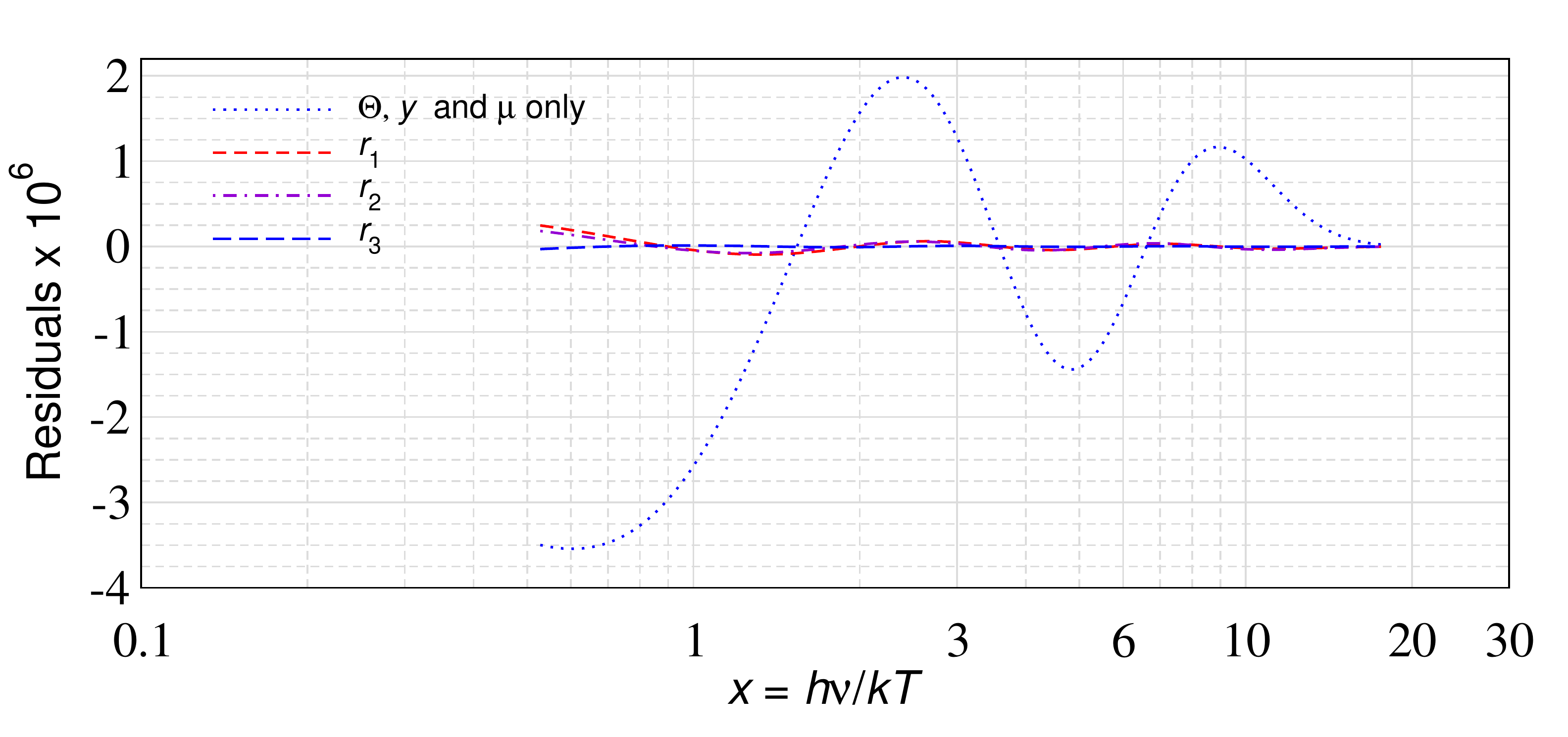}
\\
\caption{Distortion (i.e., $x^3\Delta n_x$) after a single injection at $\zh=\pot{5}{4}$ with $\Delta \rho_\gamma/\rho_\gamma=10^{-5}$ for various representations of the signal. The labels gives the maximal spectral component in the respective basis aside from the standard $\Theta, y$ and $\mu$-description. The `exact' result was obtained with the computation basis up to $Y_{15}$. The simple $\Theta$, $y$ and $\mu$-descriptions fails at the level of several tens of percent in particular at low frequencies. On the other hand, the distortion is extremely well represented once $r_2$ or $r_3$ are included. This is a compression of the information by a factor of more than $\simeq 5$.}
\label{fig:YN_convergence_res}
\end{figure}

For our analysis, we pre-compute $L$ for $N=15$ and $M=6$, but usually will only need $r_1$, $r_2$ and $r_3$ to obtain a highly accurate representation of the full $Y_k$ basis result. Even for computations of power spectra, this significantly reduces the dimensionality of the problem (see paper III). As shown in Fig.~\ref{fig:YN_convergence_res}, the residual distortion representation performs as well as the computation basis but with a lot fewer components (see discussion next Section).

\subsection{Performance and convergence}
We are now in the position to compare the performance of the {\it observation basis} in representing the distortion solutions obtained by using the computational $Y_k$ basis. Two aspects are immediately worth noting: since the observation basis has a limited frequency coverage, it will not provide a description of the distortion solution {\it outside} this domain. This is analogous to having limited sky coverage, although there the properties of the spherical harmonic basis allows for some level of statistical deconvolution in CMB analyses \citep{Hivon2002MASTER}. For the distortion spectra, this inversion problem will not be possible unless as many distortion parameters as basis parameters are observed accurately.

Second, the number of independently observable modes will depend on the frequency domain and frequency resolution as well as  the sensitivity of the experiment. For example, it has been demonstrated that distinguishing $\mu$-type distortion spectra benefits from having frequency channels below $\simeq 30\,\GHz$ \citep{abitbol_pixie, Rotti2021MILC, Rotti2022}. However, a more comprehensive exploration of these dependencies aspects is beyond the scope of this paper, and for our illustrations we will stick to the modes shown in Fig.~\ref{fig:PCAs}.

To illustrate the performance of the observation basis, we consider the distortion caused by a single energy injection at $\zh=\pot{5}{4}$ with $\Delta \rho_\gamma/\rho_\gamma=10^{-5}$. In this regime, the residual distortion contributions are expected to be largest and hence the departures from the standard $\mu$ and $y$ description are maximized. Looking at Fig.~\ref{fig:YN_convergence_res}, we can immediately see that only the first few residual distortion spectra are needed to accurately represent the distortion shape at the level of a few percent of the dominant signal. This is a significant compression of the required information for the signal processing. However, it also implies that from the precise distortion shape not as much information can be directly extracted unless a very large distortion signal is present or extremely high sensitivity is achieved \citep{Chluba2013PCA}.

\begin{figure}
\centering
\includegraphics[width=0.85\columnwidth]{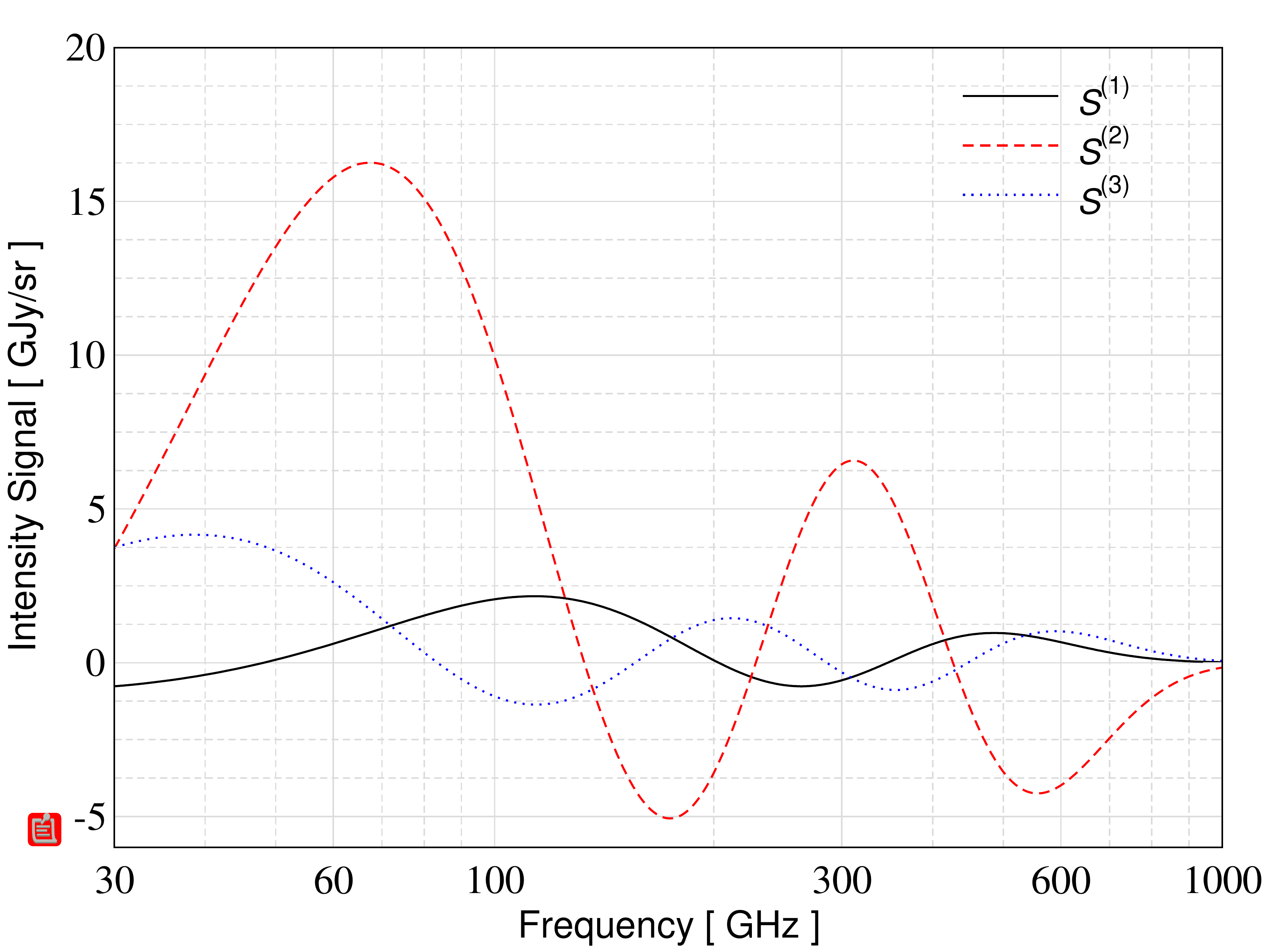}
\caption{First three residual distortion eigenmodes obtained for $\Delta \nu=1\,{\rm GHz}$ in the range $\nu_{\rm min}=30\,{\rm GHz}$ to $\nu_{\rm max}=1000\,{\rm GHz}$ and with an explicit photon number constraint to deproject $G$. These signals are orthogonal to the standard $Y$ and $M$ spectra and also among each other, but no longer leave the amplitudes of $G$ unaltered. They have all been normalized to carry an energy of $\Delta \rho_\gamma/\rho_\gamma=4$.}
\label{fig:PCAs_G}
\end{figure}

\subsection{Caveats of the observation basis and alternative description}
We would like to highlight a few important aspects of the observation basis. While by construction, $Y$, $M$ and the $Y_k$ spectra do not carry photon number, the same is not true for the residual distortion spectra. This implies that in the new representation, not only the $y$ and $\mu$ parameters change but also the temperature parameter is affected. Concretely, we have $\Theta\approx \pot{2.6}{-12}$, $y\approx\pot{6.1}{-6}$ and $\mu\approx \pot{3.0}{-8}$ for the example shown in Fig.~\ref{fig:YN_convergence_res} in the $Y_{15}$-representation. When projecting onto the observation basis we find $\Theta_{\rm o}\approx\pot{-9.2}{-7}$, $y_{\rm o}\approx\pot{1.3}{-6}$ and $\mu_{\rm o}\approx \pot{1.3}{-5}$. 
We injected energy at a redshift where DC and BR are already very inefficient, such that $\Theta$ based on scattering physics alone should be negligibly small. After the change of basis, in particular $\mu_{\rm o}$ picks up a noticeable contribution and $\Theta_{\rm o}$ even drops below zero.
This effect is known and originates from the fact that the residual distortion construction is based on intensity projections \citep[see Fig.~2 of][]{Chluba2013PCA}. The chosen procedure is most close to what would be obtained using standard component separation methods in future spectrometer analyses \citep[e.g.,][]{Rotti2021MILC, Remazeilles2022, Rotti2022}.
Although the total energetics of the problem and also the spectrum remain unchanged by the change of the representation, this behavior seems ambiguous.

\begin{figure}
\centering
\includegraphics[width=\columnwidth]{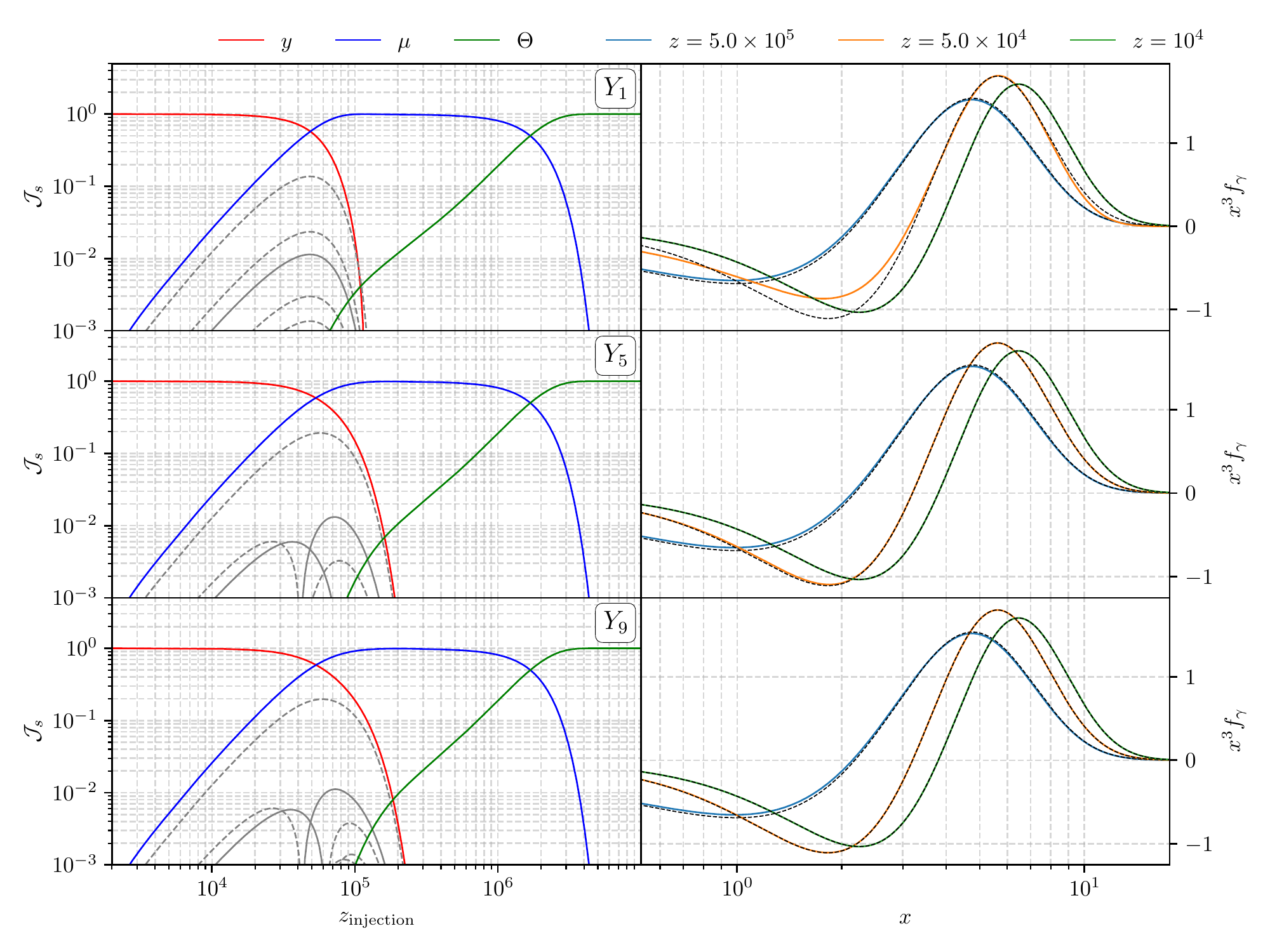}
\caption{As for Fig.~\ref{fig:branch_greens_YNmax}, but now with results cast into the ``scattering'' basis. Notice that in contrast to Fig.~\ref{fig:fig:branch_greens_obs} there is no production of a negative temperature shift, since here we enforce a strict number conservation of the residual modes, meaning no distortion shape can project onto $\Gspec$ in the change of basis. Again the spectra show no change compared to the computation basis.}
\label{fig:fig:branch_greens_scat}
\end{figure}

\begin{figure}
\centering
\includegraphics[width=0.72\columnwidth]{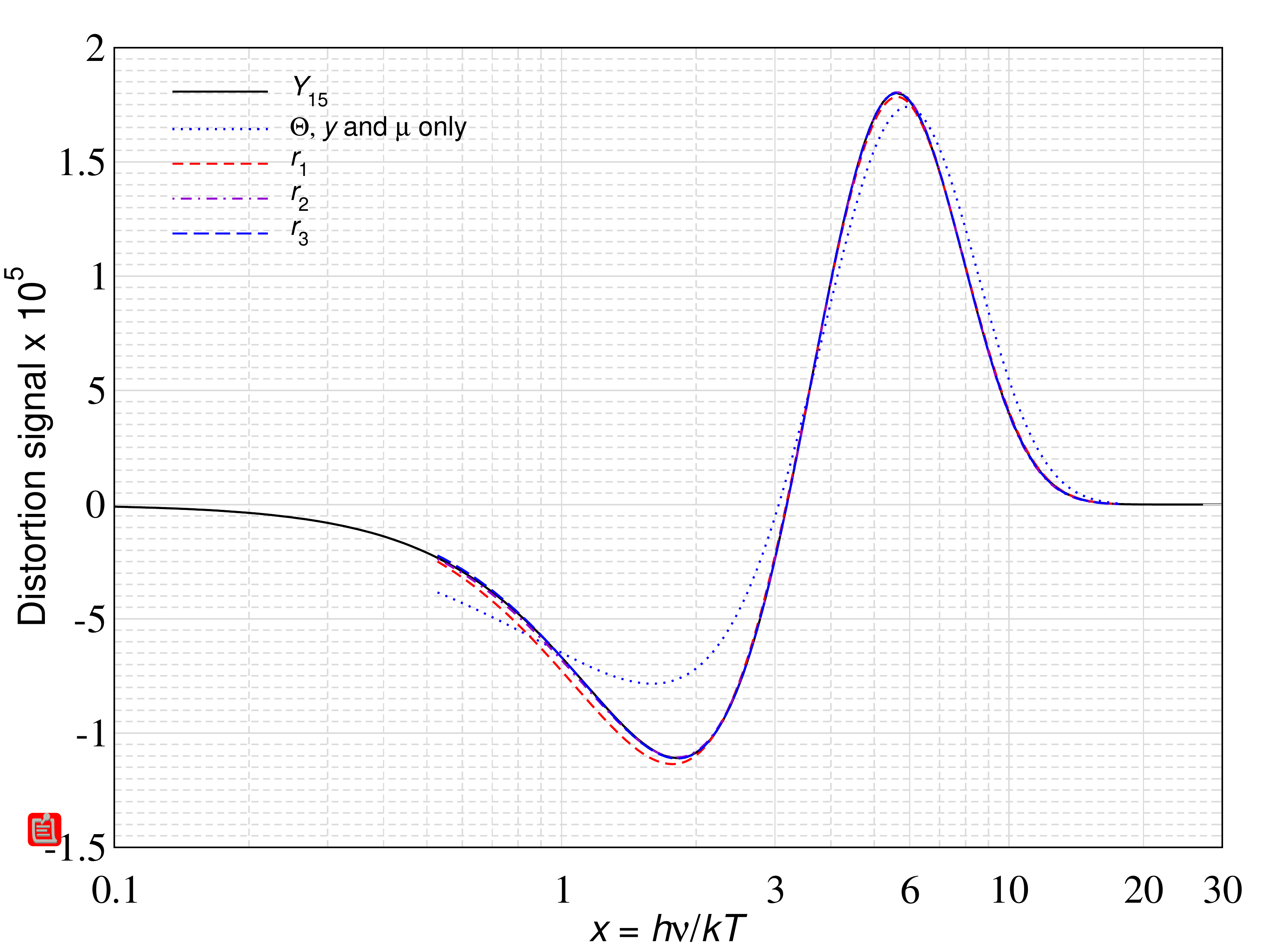}
\\
\includegraphics[width=0.72\columnwidth, trim={0cm 9cm 0 0cm},clip]{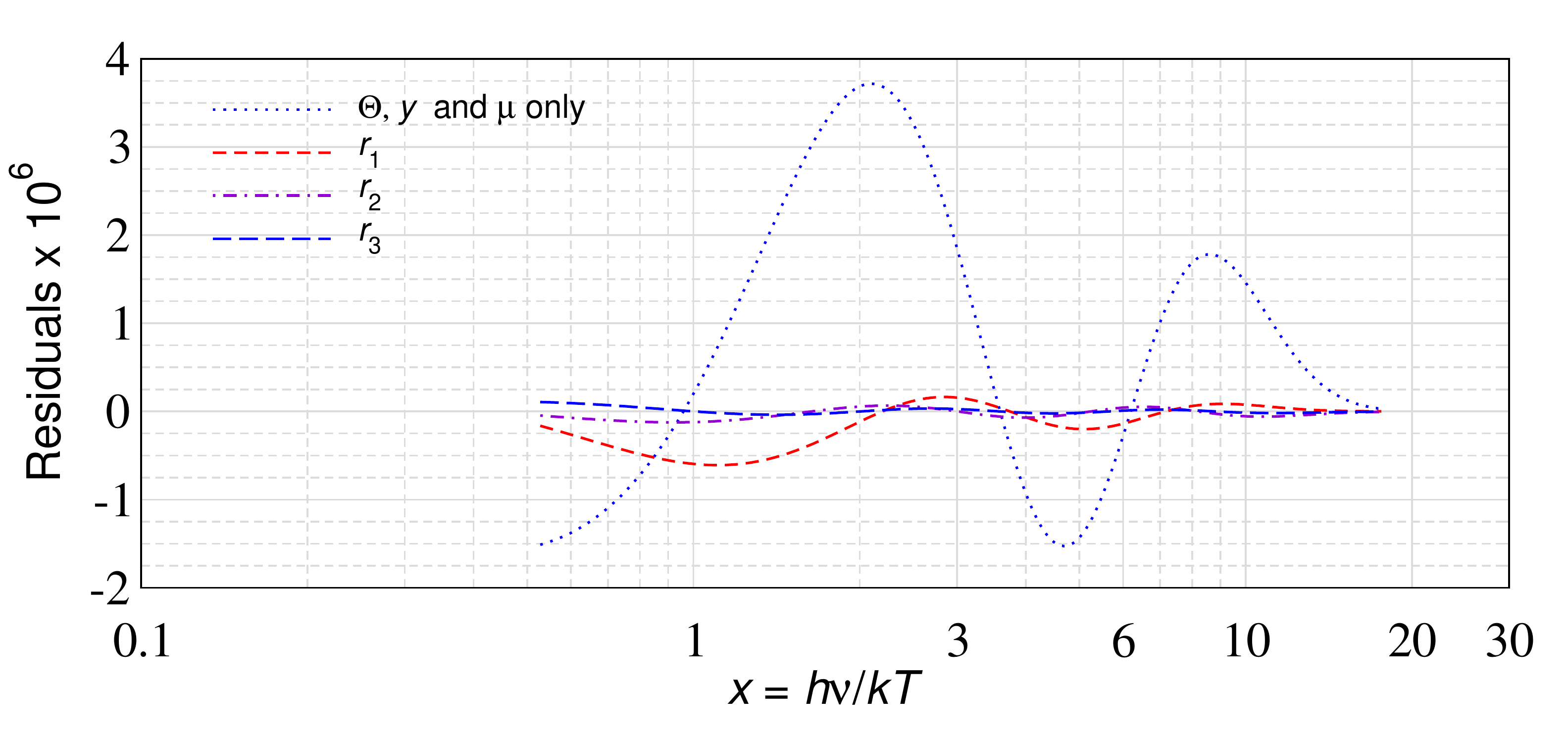}
\\
\caption{Same as Fig.~\ref{fig:YN_convergence_res} but using a photon number constraint to obtain the value for $\Theta$.}
\label{fig:YN_convergence_res_G}
\end{figure}

An alternative observational procedure, without this apparent ambiguity, could be to fix the temperature parameter $\Theta$ based on the number density of the photon field. In this case, one could fully orthogonalize $G$ to the distortion space and construct a pure residual {\it scattering} distortion representation that is unaffected by the aforementioned effects. In Fig.~\ref{fig:PCAs_G}, we show the result for the basis vectors in this alternative construction procedure. While generally very similar to the previous set of distortion modes (see Fig.~\ref{fig:PCAs}), the alternative modes show a slightly differing pattern and overall trend. These modes can only be use in cases where the temperature contribution can be independently separated, as the modes no longer are orthogonal to $G(x)$.

Fig.~\ref{fig:fig:branch_greens_scat} shows how this basis again stabilises the $y$ and $\mu$ amplitudes across basis size while reproducing the same spectrum as the other bases. The difference however is that now the first residual mode constitutes a more dominant fraction of the energy, and the temperature shift never takes on its effective negative value. This is closer to the full scattering physics, since now number is conserved, but the description is not akin to a realistic observation of the sky. In paper III we will use these two bases wherever they are most illustrative, but always being careful and explicit.

In Fig.~\ref{fig:YN_convergence_res_G}, we demonstrate that these alternative modes also represent the distortion shape very well, with the difference being that the contribution from $G$ was fixed independently using a number density constraint. In the $Y_{15}$ representation, one has $\Theta\approx \pot{2.6}{-12}$, $y\approx\pot{6.1}{-6}$ and $\mu\approx \pot{3.0}{-8}$ as before. Taking the full spectrum and imposing the photon number constraint to obtain the amplitude of $G$ and then fitting for $y$ and $\mu$, we obtain $\Theta\approx \pot{2.6}{-12}$, $y\approx\pot{1.6}{-6}$ and $\mu\approx \pot{7.7}{-6}$. Just like before, we see a significant change in the values for $y$ and $\mu$, but 
this time no change to $\Theta$. The total energy carried by $G$, $Y$ and $M$ is $\Delta \rho_\gamma/\rho_\gamma\simeq 4y+\mu/1.4\approx \pot{1.2}{-5}$, implying that the residual distortion contributes $\Delta \rho_\gamma/\rho_\gamma\simeq -\pot{0.2}{-5}$. In contrast, for the $Y_{15}$ representation we have $\Delta \rho_\gamma/\rho_\gamma\approx \pot{2.5}{-5}$ stored in the $G$, $Y$ and $M$ components, implying that about $\Delta \rho_\gamma/\rho_\gamma\approx -\pot{1.5}{-5}$ is in the $Y_{k>0}$ terms, which is {\it no} small total correction. If we compare all this to the lowest order computation using only $\Theta$, $y$ and $\mu$ in the ODE (i.e., a $Y_0$-representation) we obtain $\Theta\approx \pot{1.1}{-9}$, $y\approx\pot{1.6}{-6}$ and $\mu\approx \pot{5.1}{-6}$. This demonstrates that the $\mu$ and $y$ decomposition is well captured by the alternative distortion eigenmodes.

However, in the outlined alternative procedure an observer has to evaluate the number integral $\propto \int x^2 \Delta n_x \id x$ of the photon field, which in experimental settings has several challenges. First, unless the distortion is measured in a sufficiently wide range of frequencies, this number integral would not evaluate accurately. Specifically, even $\int x^2 Y\id x$ and $\int x^2 M\id x$ are no longer guaranteed to vanish, thereby breaking the `photon number orthogonality'. Second, to carry out the integral, the frequency sampling has to be fine, which again usually runs into observational difficulties. Third, the estimation of errors will be non-standard since the observable is based on weighted sums of fluxes. Therefore, this approach is not expected to be realized in actual observations. Nevertheless, for theoretical calculations, we can use it for illustration when the focus is on the energetics of the problem.  
We will therefore refer to this alternative basis as {\it scattering basis}, given that it is constructed to focus on the spectral shapes that are introduced purely by Compton scattering terms, which conserve photon number.
We further discuss the benefits and differences of changing the basis in paper III.

\section{Discussion and conclusions}
In this work we obtained an approximate ODE treatment for the thermalisation Green's function, which captures most aspects of the full calculation using an extended spectral basis to describe the residual distortion evolution (Sect.~\ref{sec:ODE_Greens}).
Instead of the expensive `top-hat' frequency binning we use a spectral basis that is derived from boosts of the $y$-distortion spectrum. 
This reduces the computational burden by a factor of $\simeq 10^3$, thereby providing one of the main steps towards formulating a generalised photon Boltzmann hierarchy, that will allow us to compute the evolution of distortion anisotropies at first order in perturbation theory (see papers II and III).
We also clarify how the computational spectral basis can be compressed into fewer distortion shapes that can be distinguished with a given experimental configuration, introducing the {\it observation} and {\it scattering} basis (Sect.~\ref{sec:observables}). 

The new ODE representation of the thermalisation Green's function given here is not perfect, because not all the spectral shapes can be spanned by the basis functions we choose (see Fig.~\ref{fig:YN_bad_convergence}). However, we have demonstrated that this is not a severe problem for the average distortion evolution, which specifically relies on conversion of $Y(x)$ [as the main distortion source] to $M(x)$ and $G(x)$.
The representation of the full Green's function could probably be improved by studying the eigenfunctions of the Kompaneets and boost operators more carefully. In addition, weighting schemes and a modified truncation of the distortion basis could likely improve the performance. One could also refine the treatment of photon emission processes, including the effect of the new distortion shapes. This is expected to modify the thermalisation efficiency, a problem that may be solved perturbatively. 
Nevertheless, the novel ODE representation of the average thermalisation Green's function is sufficiently accurate for approximate applications to SD anisotropies, as we show in papers II and III.

Overall, this paper is the first in a series of works discussing the evolution of SD anisotropies generated by various physical mechanisms and how these might be constrained with future CMB spectrometers and imagers. The results from these works should open the path for more realististic SD anisotropy forecasts over a wide range of physics which previously were not possible. This will hopefully spur additional activity on CMB spectral distortions, uniting the efforts of CMB imaging and spectrometer approaches for probing the early Universe.

{\small
\section*{Acknowledgments}
We thank Eiichiro Komatsu, Aditya Rotti and Rashid Sunyaev for stimulating discussion.
We furthermore thank Nicola Bartolo, Richard Battye, Daniele Bertacca, William Coulton, Bryce Cyr, Colin Hill, Antony Lewis, Rishi Khatri, Sabino Matarrese, Atsuhisa Ota, Enrico Pajer and Nils Sch\"oneberg for comments on the manuscript.
This work was supported by the ERC Consolidator Grant {\it CMBSPEC} (No.~725456).
TK was also supported by STFC grant ST/T506291/1.
JC was furthermore supported by the Royal Society as a Royal Society University Research Fellow at the University of Manchester, UK (No.~URF/R/191023).
AR acknowledges support by the project "Combining Cosmic Microwave Background and Large Scale Structure data: an Integrated Approach for Addressing Fundamental Questions in Cosmology", funded by the MIUR Progetti di Ricerca di Rilevante Interesse Nazionale (PRIN) Bando 2017 - grant 2017YJYZAH.
}


{\small
\bibliographystyle{JHEP}
\bibliography{bibliography,Lit}
}

\appendix

\section{Useful operator properties}
\label{app:operator_props}
We can somewhat reduce the complexity of the above calculations by studying the properties of -- and relationships between -- the operators $\DiffO$, $\DiffstarO$ and $\boostO$. This will furthermore illustrate the suitability of the expanded basis $\{Y_k(x)\}$.

We first note that two of the main operators commute with one another:
\begin{equation}
    \big[\DiffO, \boostO\big] = 0,
\end{equation}
thus implying the potential existence of a shared eigenbasis. A logical step is to express the \textit{larger} of the operators in terms of the other, revealing the identities
\begin{equation}
\label{eq:diff_properties}
    \DiffO = \boostO(\boostO-3)=(\boostO-3) \boostO =-3\boostO+ (\boostO)^2 = 4\boost + x^2\,\partial_x^2.
\end{equation}
The final equality can be easily found with a single application of the chain rule. However, it hints towards a more generic recurrence relation, which yields the following combinatoric sum:
\begin{equation}
    (\boostO)^k = (-1)^k\sum_{m=1}^k \sterling{k}{m} \,x^m\partial_x^m,
\end{equation}
where the square brackets indicate \textit{Stirling set numbers}, which counts partitions of an $k$-set into $m$ nonempty subsets. This expansion of the boost operator reveals that the reverse operation is non-trivial -- very specific weighted sums of $x^m\partial_x^m$ terms are needed to make a boost operator with some power. Because of this, it is useful to be able to compose these expanded $\boostO$ terms directly:
\begin{equation}
    x^a\,\partial_x^a \, x^b\,\partial_x^b = 
    \sum_{k=0}^{a}\frac{b!}{(b-a+k)!}\,\binom{a}{k}\,x^{b+k}\,\partial_x^{b+k}.
\end{equation}
As noted in \cite{Chluba:eulerian_numbers}, $x^k\partial_x^k\nbb$ has a recursion relation allowing for another combinatoric analytic solution
\begin{equation}
    x^k\partial_x^k\nbb = \frac{(-x)^k \expf{-x}}{(1-\expf{-x})^{k+1}}\sum_{m=0}^{k-1}\,\eulerian{k}{m}\, \expf{-mx}  \hspace{1cm} (k > 0),
\end{equation}
where the angle brackets denote \textit{Eulerian numbers}, defined as the number of permutations of the numbers $1$ to $m$ in which exactly $k$ elements are greater than the previous element. This expression has very good convergence properties when summed starting from the highest power of $\expf{-mx}$.

Defining $\mathcal{H}_k(x)=(-x)^k \expf{-x}/(1-\expf{-x})^{k+1}$, we are now in a position to write general expressions using the above formulae: 
\begin{gather}
    \DiffO^N = \sum_{k=0}^N \,\binom{N}{k}\,3^k \sum_{m=1}^{2N-k} \,\sterling{2N-k}{m}\, x^m\,\partial_x^m, \\
    (\boostO)^N\nbb = (-1)^N\sum_{k=1}^N \,\sterling{N}{k}\, \mathcal{H}_k(x)\sum_{m=0}^{k-1}\,\eulerian{k}{m}\, \expf{-mx} \hspace{1cm} (N > 0),\\
    \DiffO^N(\boostO)^M \nbb = (-1)^M\sum_{k=0}^N \,\binom{N}{k}\, 3^k \sum_{\ell=1}^{2N+M-k}\,\sterling{2N+M-k}{\ell}\, \mathcal{H}_\ell
    \sum_{m=0}^{\ell-1}\,\sterling{l}{m}\,\expf{-mx}.
\end{gather}
We can generate the basis functions $Y_N$ from the above expressions noticing that according to our convention $Y_N=(\boostO/4)^N \Yspec = \DiffO(\boostO/4)^N \nbb$: 
\begin{equation}
    Y_N = (-1/4)^N\sum_{k=1}^{N+2} \left( 3\sterling{N+1}{k} + \sterling{N+2}{k} \right) \mathcal{H}_k \sum_{m=0}^{k-1}\,\eulerian{k}{m}\, \expf{-mx}.
\end{equation}
Note that we have used $\sterling{N+1}{N}=0$ to simplify the above expression, bringing two different powers of derivatives under a single summation sign. Below we provide a few examples:
\begin{align}
&
\begin{aligned}
\Ynspec{1}(x) =
\frac{\expf{-x} \left(\expf{-2 x}+4 \expf{-x}+1\right) x^3}{4\left(1-\expf{-x}\right)^4}
-\frac{3\expf{-x}\left(\expf{-x}+1\right) x^2}{2\left(1-\expf{-x}\right)^3}
+\frac{\expf{-x} x}{\left(1-\expf{-x}\right)^2},
\end{aligned}\\
&\nonumber\\&
\begin{aligned}
\Ynspec{2}(x) = 
&\frac{\expf{-x} \left(\expf{-3 x}+11 \expf{-2 x}+11 \expf{-x}+1\right) x^4}{16\left(1-\expf{-x}\right)^5}
-\frac{9\expf{-x} \left(\expf{-2 x}+4 \expf{-x}+1\right) x^3}{16\left(1-\expf{-x}\right)^4}\\
&\qquad+\frac{ \expf{-x}\left(\expf{-x}+1\right) x^2}{\left(1-\expf{-x}\right)^3}
-\frac{\expf{-x} x}{4\left(1-\expf{-x}\right)^2},\\
\end{aligned}\\
&\nonumber\\&
\begin{aligned}
\Ynspec{3}(x) = 
&\frac{\expf{-x} \left(\expf{-4 x}+26 \expf{-3 x}+66 \expf{-2 x}+26 \expf{-x}+1\right)x^5}{64\left(1-\expf{-x}\right)^6}
-\frac{13 \expf{-x} \left(\expf{-3 x}+11 \expf{-2 x}+11 \expf{-x}+1\right)x^4}{64\left(1-\expf{-x}\right)^5}\\
&\qquad+\frac{43 \expf{-x} \left(\expf{-2 x}+4 \expf{-x}+1\right)x^3}{64\left(1-\expf{-x}\right)^4}
-\frac{9 \expf{-x} \left(\expf{-x}+1\right)x^2}{16\left(1-\expf{-x}\right)^3}
+\frac{\expf{-x} x}{16\left(1-\expf{-x}\right)^2}.
\end{aligned}
\end{align}
Despite the progress made above, the overall problem is not fully closed via combinatoric sums. The operator $\diffusionStar\in\KompO$ does not commute with the others. Instead we find
\begin{gather}
    \left[\frac{1}{x}\DiffstarO, \boostO\right] = 0, \qquad
    \left[\DiffstarO \frac{1}{x}, \boostO\right] = 0, \qquad \left[\DiffO, \frac{1}{x}\DiffstarO\right] = 0, \qquad \left[\diffusion, \DiffstarO\frac{1}{x}\right] = 0,
\end{gather}
showing that no such shared basis will exist, and thus for now we resort to the approximate numerical projections discussed in the main text (see especially Fig.~\ref{fig:KY_approx}).

However, some more progress can be made my realising that $\DiffstarO$ always appears in conjunction with with the factor $A=(1+2 n_{\rm bb})$. It can be shown that $A=\frac{1}{x}\left(4 + \frac{\Yspec}{\Gspec}\right)$. This loose factor of $1/x$ combines nicely with the commutators noted above. Specifically we can then write
\begin{equation}
    \DiffstarO A = -(\boostO-3)\left( 4+\frac{\Yspec}{\Gspec}\right).
\end{equation}
Combining this with Eq.\eqref{eq:diff_properties} we can write
\begin{equation}
    \KompO = (\boostO-3)\left[ \boostO -4 -\frac{\Yspec}{\Gspec} \right] = \boostO^2 -7\boostO+ 12 +3\frac{\Yspec}{\Gspec} -\boostO\frac{\Yspec}{\Gspec},
\end{equation}
which essentially distils the \textit{misbehaving} part of the Kompaneets operator to the previously named $y$-weight factor $w_y= \Yspec/\Gspec$.

This expression of the Kompaneets operator makes it clearer to see how certain results arise algebraically. Consider for example that $\KompO \Gspec=-\Yspec$, and similarly $\KompO \Mspec=-\Yspec$, where the latter result follows from the former together with $\KompO (\Gspec/x)=0$. The spectral shape, $Y_1$, appears as intermediate step in these calculations, but ends up cancelling. These results may not be interesting in isolation, but they emphasise the fact that the cancellations only occur for simple shapes. Once you apply $\KompO$ to a distortion shape like $\Yspec$ you naturally get $Y_1$ and $Y_2$ that do not analytically cancel.

\section{Alternative derivation of the ODE system}
\label{app:formal}
To obtain the ODE system for the evolution of the spectrum, we can also directly project the evolution equation. Making the Ansatz $\Delta n\approx \vek{B}\cdot \vek{y}$ (with definitions as in the main section for a given basis) and then inserting this into the evolution equation, Eq.~\eqref{eq:Main_Eq}, we have
\begin{align}
\label{eq:Main_Eq_muyyk}
\Theta'\Gspec(x)+\sum_{k=0}^N y'_k Y_k(x)+\mu'\Mspec(x)
&=\Theta_{\rm e}\,\Yspec(x)- \Theta\,\Yspec(x)+\sum_{k=0}^N y_k\,K_{Y_k}(x)-\mu\,\eta_M \Yspec(x).
\end{align}
Here, $\Ynspec{0}\equiv Y$, $K_{Y_k}=\KompO Y_k$ and we used the identities in Eq.~\eqref{eq:KG_KM}. Since only $\Gspec(x)$ carries number we immediately obtain $\Theta'=0$ by carrying out the number integral $\int x^2 \id x$ over this equation. Since we know that $- \Theta\,\Yspec(x)$ on the right hand side of Eq.~\eqref{eq:Main_Eq_muyyk} cancels the corresponding term in the Compton equilibrium temperature
\begin{align}
\label{eq:Compton_eq_muy}
\Theta_{\rm e}
&\approx \left(\frac{\int x^3 w_y(x)\,\vek{B} \id x}{4 E_{\nbb}}\right) \cdot \vek{y}
\approx \Theta+\sum_{k=0}^N\,\eta_{Y_k}\,y_k+\eta_M\,\mu,
\end{align}
and because there also is no term $\propto \Gspec(x)$, we only have to worry about the reduced problem
\begin{align}
\label{eq:Main_Eq_muyyk_red}
\sum_{k=0}^N y'_k Y_k(x)+\mu'\Mspec(x)
&=\left(\Theta_{\rm e}-\Theta-\eta_M\,\mu \right)\Yspec(x) +\sum_{k=0}^N y_k\,K_{Y_k}(x).
\end{align}
By performing the projections onto all function of the representation basis $\vek{R}=(Y,\Ynspec{1},\ldots,\Ynspec{k},M)^T$, we obtain the system
\begin{align}
\label{eq:Main_Eq_muyyk_red_vec}
M_R\, \vek{y}'
&=\left(\Theta_{\rm e}-\Theta-\mu\,\eta_M\right)\,\vek{b}_Y+K\, \vek{y}.
\end{align}
where $\vek{y}=(y,y_1,\ldots,y_k,\mu)^{T}$ and $M_{R,ij}=\langle R_i | R_j \rangle$ is the full mixing matrix. We also have the source vector $b_{Y,i}=\langle R_i|R_0\rangle=\langle R_i|Y\rangle$ and Kompaneets matrix $K_{ij}=\langle R_i|K_{Y_j}\rangle$. 

As already explained in the main text, the system above will not yield a solution that correctly conserves energy (although it will become better and better the more $Y_k$ are included). We therefore replace the last row in the matrices $M_R$ and $K$ and the last entry in $\vek{b}_{Y}$ with the corresponding energy equation (as shown in the main text). The modified system has the same form as Eq.~\eqref{eq:Main_Eq_muyyk_red_vec}, just with redefined matrices and vectors which we do not explicitly distinguish in the notation. The system can be solved for $\vek{y}'$ to obtain the evolution equation for $y, y_1, \ldots, y_N, \mu$ as
\begin{align}
\label{eq:Main_Eq_muyyk_red_vec_mod_sol}
\vek{y}'
&=\left(\Theta_{\rm e}-\Theta-\mu \eta_M\right)\,M_R^{-1}\, \vek{b}_Y+M_R^{-1}\,K\, \vek{y}.
\end{align}
The rows of the matrix ${M}_R^{-1}\,{K}$ are composed of the representation vectors for the operators $K_{Y_k}$. Note that the matrix ${K}$ is an $(N+2)\times (N+1)$ matrix, while ${M_R}^{-1}$ is an $(N+2)\times (N+2)$ matrix, such that $M_R^{-1}\,{K}$ also is an $(N+2)\times (N+1)$ matrix. In addition we have $M_R^{-1}\, \vek{b}_Y=\delta_{i0}$, which simply follows from the fact that $\tilde{\vek{b}}_Y$ is the first column vector of the matrix ${M_R}$. Since the matrix $M_R^{-1}\,{K}$ can be determined by independently solving for the representations of $K_{Y_k}$ in terms of the representation basis $\vek{R}$, this means we have proven the equivalence with the approach used in the main text.

\end{document}